\documentclass[a4paper,11pt]{article}
\pdfoutput=1 % if your are submitting a pdflatex (i.e. if you have
\usepackage{jcappub} % for details on the use of the package, please
\usepackage{url}	% Advanced maths commands
\usepackage{amsfonts,amssymb,amsmath}

\usepackage{graphicx}
\usepackage{subfigure}
\usepackage{color}
\usepackage{multirow}
\usepackage{multicol}
\usepackage{tikz}
\usepackage{float} %added by Shabbir, for fig placement
\usepackage{aas_macros}
\usepackage[normalem]{ulem}
\usepackage{array}
\usepackage{comment}
\usepackage{appendix}
\DeclareMathAlphabet{\mathsc}{OT1}{cmr}{m}{sc}
\newcommand{\QHII}{\ensuremath{Q_\mathsc {HII}}}
\newcommand{\da}[1]{{ #1}}

\title{Model-independent Reconstruction of UV Luminosity Function and Reionization Epoch}
\author[a,b,c,d]{Debabrata Adak,}
\author[c,d,e]{Dhiraj Kumar Hazra,}
\author[f]{Sourav Mitra,}
\author[c,g]{Aditi Krishak}
\affiliation[a]{Instituto de Astrofísica de Canarias, E-38200 La Laguna, Tenerife, Spain}
\affiliation[b]{Departamento de Astrofísica, Universidad de La Laguna, E-38206 La Laguna, Tenerife, Spain}
\affiliation[c]{The Institute of Mathematical Sciences, CIT Campus, Chennai 600113, India}
\affiliation[d]{Homi Bhabha National Institute, Training School Complex, Anushakti Nagar, Mumbai 400085, India}
\affiliation[e]{INAF/OAS Bologna, Osservatorio di Astrofisica e Scienza dello Spazio di Bologna, via Gobetti 101, I-40129 Bologna, Italy}
\affiliation[f]{Department of Physics, Surendranath College, 24/2 M. G. Road, Kolkata 700009, India}
\affiliation[g]{Department of Physics and Astronomy, University of Southern California, University Park, Los Angeles, CA 90089, USA}
\emailAdd{adak@iac.es, dhiraj@imsc.res.in, hisourav@gmail.com, krishak@usc.edu}

\abstract{
    We conduct a first comprehensive study of the Luminosity Function (LF) using a non-parametric approach. We use Gaussian Process to fit available luminosity data between redshifts $z \sim 2-8$.  Our free-form LF in the non-parametric approach rules out the conventional Schechter function model to describe the abundance-magnitude relation at redshifts $z=3$ and $4$. Hints of deviation from the Schechter function are also noticed at redshifts 2, 7 and 8 at lower statistical significance. Significant deviation starts for brighter ionizing sources at $M_{\rm UV} \lesssim -21$. The UV luminosity density data at different redshifts are then derived by integrating the LFs obtained from both  methods with a truncation magnitude of $-17$. In our analysis, we also include the first 90 arcmin$^2$ JWST/NIRCam data at $z \sim 9-12$. Since at larger magnitudes, we do not find major deviations from the Schechter function, the integrated luminosity density differs marginally between the two methods. Finally, we obtain the history of reionization from a joint analysis of UV luminosity density data along with the ionization fraction data and Planck observation of Cosmic Microwave Background. The history of reionization is not affected by the deviation of LFs from Schechter function at lower magnitudes. We derive reionization optical depth to be $\tau_{\rm re}=0.0494^{+0.0007}_{-0.0006}$ and the duration between 10\% and 90\% completion of ionization process is found to be $\Delta z\sim 1.627^{+0.059}_{-0.071}$.} 
    
\begin{document}
\maketitle
\flushbottom

\section{Introduction}\label{sec:intro}
The process of cosmic reionization is an outstanding problem in extragalactic astronomy. During the redshifts $z \sim 6-20$ almost all of the hydrogen in the universe became ionized~\citep{Fan:2006,McGreer:2015,Hazra:2017,Planck-VI:2020,paoletti2024asymmetry}. A similar process was followed for helium at a later time and helium reionization was completed at around $z\sim 2.5 - 3.5$ \cite{2016ApJ...825..144W}. For this process, the high-redshift ($z \gtrsim 6$) star-forming galaxies are often considered to be the dominant contributors of ionizing photons since the abundance of quasars dramatically declines at $z \sim 6$~\citep{Fontana:2010, Willott:2010,Onoue_2017}. A variety of theoretical and observational studies have shown that the contribution of Active Galactic  Nuclei has a very minimal ($\lesssim$ 1\%) contribution to the total ionisation budget at $z\geq$ 6 \citep{10.1093/mnras/stx2194,dayal:2020}. Most of the studies suggest that star-forming galaxies inside the low mass halo ($M_h \lesssim 10^{9.5} M_{\odot}$) are sufficient enough to complete the reionization process \cite{10.1093/mnras/stw1015, TR_Choudhury:2005}. Therefore, their time-dependent abundance, and the redshift evolution of UV luminosity density ($\rho_{\rm UV}$) derived from rest-frame UV luminosity function (LF) of galaxies are of significant interest for understanding reionization history (see,~\cite{Robertson:2010}).  

Significant advancement has been made in determining the behaviour of LFs at redshifts $z \sim 4-10$ using Hubble Frontier Fields (HFF, \cite{Lotz:2017}) survey data up to magnitude $\simeq- 15$~\citep{Schenker_2013,Ellis:2013,McLure:2013}. More recently, the same has been studied in~\cite{Ishigaki:2015, Ishigaki:2018} and~\cite{Bouwens:2015} with galaxy candidates down to magnitude $\simeq-22$. However, along with studying the faint end of the UV LF, it is also important to investigate the bright-end shape of LF. While the conventional Schechter function \citep{Schechter:1976} seems to be a good description of the LFs for ionizing candidates at the fainter end of the magnitude, an exponential decline has been reported in their number densities at the bright-end \citep{Loveday:2012}. It is thought to be caused by heating from an active galactic nucleus (AGN, \cite{Bower:2006}), inefficient cooling of gas inside high-mass dark matter halos at low redshifts \citep{Benson:2003} etc. The studies in \cite{Ono:2018, Stevans:2018} and \cite{Adams:2020}  reported an excess of number density compared to that described by the Schechter best fit at $z \sim 4-7$. 

 With the improvement of signal-to-noise ratio in astrophysical observations, search for beyond standard model has usually taken two approaches, namely, model building and model independent reconstructions. In this paper, for the first time, we search for the deviations beyond Schechter function using model independent approach. Model building approaches follow either phenomenological or theoretical constructions. Recent works of \cite{Ono:2018} and \cite{Harikane:2022} found that both double power-law (DPL, \cite{Bowler:2012}) and lensed Schechter function \citep{Wyithe:2011} are better fit to the UV luminosity data. Their studies include  brightest galaxy candidates observed in the early data from James Webb Space Telescope (JWST, \cite{Rigby:2023}) and Great Optically Luminous Dropout Research data of Subaru HSC (GOLDRUSH) (corrected for AGN contribution) respectively. Publications~\cite{Stefanon:2019} and \cite{bowler20} independently established that LFs are more consistent with DPL at $z \sim 9-10$.  \da{DPL model is a simple modification of Schechter function with introduction of another power law (see Eq.1 of \cite{Bowler:2012}) instead of exponential decrease of Schechter form after a certain magnitude.} \da{ The lensed Schechter function is a convolution of magniﬁcation effect on the observed shape of
the galaxy UV luminosity function determined by intrinsic Schechter function. This magnification bias is determined by strong-lensing optical depth that is fraction of strong-lensed random lines of sight. In current literature \citep{Harikane:2022} the amount of modification is based on upper limit of optical depth
by foreground sources estimated in \cite{Takahashi_2011,2015MNRAS.450.1224B} along with pre-determined Schechter parameters. These parameterization allows us to find any excess or drop in abundance at a certain end of magnitude. However, these approaches restrict the flexibility to determine the actual trend of luminosity-magnitude relation throughout the scale of magnitude. Model independent approach here helps in highlighting the significance of the outliers and the hints for model building. The Gaussian process (GP) is a Bayesian model independent reconstruction that provides the possible departure from the Schechter function throughout the scale of the magnitude. The GP hyperparameters directly highlight the significance with which the data disfavors the Schechter function. It also finds the short range or long range nature of the deviation of LF from parametric models.}

In this paper, we address the constraints on reionization history in a two-step process. First, we reconstruct the profile of the LFs using GP, a nonparametric \da{Bayesian} method at redshifts $z \sim 2-8$, and Schechter function model at redshifts  $z \sim 2-12$ using the HFF data, early JWST data and HSC data. We compare these model-independent LFs with the \da{best fit Schechter function model}. We derive the UV luminosity densities by integrating LFs, and investigate the redshift evolution of UV luminosity densities derived from both profiles. Finally, we use these luminosity densities to constrain the history of reionization along with jointly fitting two other observational data: the CMB power spectrum data of temperature and polarization anisotropy from Planck observation~\citep{Planck-VI:2020} and neutral hydrogen fraction data from galaxy, quasar and gamma-ray burst observations.

The paper is organised as follows. In~\autoref{sec:methods} we briefly discuss the reionization process, different functions used and details of Gaussian process regression. In~\autoref{sec:data} we describe the ancillary data sets used in this work. In~\autoref{sec:results} we present our results for the UV luminosity functions obtained using two different methods, their comparison, derived UV luminosity densities and the corresponding constraints on reionization history. Finally in~\autoref{sec:summary} we summarise the work.  

\section{Methodology}\label{sec:methods}
\subsection{Cosmic Reionization}
The process of reionization of the intergalactic medium (IGM) is a balance between ionization of hydrogen and helium atoms by cosmic photons and recombination of free electrons and protons to form neutral hydrogen and helium. Analytical and numerical modelling of this process traces a long history \citep{Barkana_Lobe:2001,TR_Choudhury:2005,Kuhle:2012} (see the references therein). The process is studied by the redshift evolution of volume filling factor $\QHII$\ which is governed by the ionization histories of both hydrogen and helium. The time evolution of \QHII\ is obtained by solving the ordinary differential equation (e.g. \cite{Robertson_2013})
\begin{equation}
    \label{eq:ionise_eq}
    \ensuremath{\dot{Q}_\mathsc {HII}} = \frac{\dot{n}_{ion}}{\langle n_{H} \rangle} - \frac{\QHII}{t_{rec}},
\end{equation}
where $\langle n_{H} \rangle = \frac{X_{p}\Omega_{b}\rho_{c}}{m_{H}}$ is the mean comoving number density of hydrogen and depends on the primordial mass-fraction of hydrogen $X_p$, critical density $\rho_c$, baryon density $\Omega_b$ and $m_H$ is the mass of atomic hydrogen. $t_{rec}$ denotes the average recombination time in the IGM, 
\begin{equation}\label{eq:rec_time}
    t_{rec} = \frac{1}{C_{\rm HII}\alpha_{B}(T)\left(1 + \frac{Y_p}{4X_p}\right)\langle n_{H} \rangle (1 + z)^3},
\end{equation}
 where $\alpha_{B}(T)$ is the recombination coefficient for hydrogen (we assume the IGM temperature T to be 20,000 K) and $Y_p = 1 - X_p$ is the primordial helium abundance. The clumping factor $C_{\rm HII}$ accounts for the inhomogeneity of the IGM, and is not very well constrained from observations. Recent simulations suggest a possible range of $C_{\rm HII}$ value from 1 to 6 \citep{Iliev:2006, Pawlik:2009,Finlator:2012,Schroeder:2013}. In this work we use the fixed value of $C_{\rm HII} = 3$ \citep{Michael_Shull_2012} for simplicity. %In this work we use the following functional form derived using N-body simulations by \cite{Michael_Shull_2012}:
%\begin{equation}
%\label{eq:chii}
%    C_{\rm HII} = 2.9 \times \big[ \frac{1 + z}{6}\big]^{-1.1}.
%\end{equation}
The comoving production rate of available ionizing photons in the IGM at some redshift is 
\begin{align}
    \dot{n}_{ion} &= \int_{-\infty}^{M_{\rm trunc}} f_{esc}(M)\xi_{ion}(M)\Phi(M)L(M)dM\nonumber\\
    &=\langle f_{esc}\xi_{ion} \rangle \rho_{\rm UV}.
\end{align}
This depends on the intrinsic production rate of Lyman continuum (LyC) photons supplied by stellar populations of galaxies, which is parameterized by a numerical factor $\xi_{ion}$ to count the ionizing photons per unit UV luminosity, the escape fraction $f_{esc}$ and total luminosity density $\rho_{\rm UV}$ from star-forming galaxies with a truncation absolute magnitude $M_{\rm trunc}$. $f_{esc}$ is a crucial parameter and is not well-constrained from direct observations of LyC photons \citep{Robertson:2022}, which is mainly limited to $z\sim$ 2 -- 4.5 due to a dramatic increase in the opacity of IGM at high redshifts making direct observation of LyC photons difficult \citep{Inoue:2014}. Moreover, the trend of $f_{esc}$ with halo mass is not well understood from theoretical modelling \citep{Fernandez:2011,Mason:2018,Ma:2015}. All pre-JWST literatures suggest that a low ionisation efficiency of around $\log_{10} \xi_{ion} = 25.2$ Hz $erg^{-1}$  and hence $f_{esc}$ of 0.2 \cite{Robertson_2013} that are sufficient to give reionization history consistent with CMB data. However new JWST data suggests a higher $\xi_{ion}$ \cite{Atek:2024}  that increases at redshifts higher than $z\sim$ 9 \cite{Simmonds:2024, MU:2024}. It is apparent that $\xi_{ion}$ and $f_{esc}$ are completely degenerate parameters. A recent data-driven model-independent study \citep{Mitra:2023} found that a constant value of $f_{esc}$ for $z \geq$ 6 is permitted. Therefore, we take $f_{esc}$ as a constant factor and, instead of considering $\xi_{ion}$ and $f_{esc}$ to be independent parameters, we consider the magnitude-averaged value of $\langle f_{esc}\xi_{ion} \rangle$ as a single parameter in this work. 

Luminosity density $\rho_{\rm UV}$ is obtained from the luminosity function $\phi(M_{\rm UV})$ via integration,
\begin{equation}\label{Eq:l_density}
    \rho_{\rm UV} = \int_{-\infty}^{M_{\rm trunc}} \Phi(M)L(M)dM,
\end{equation}
where $L(M)$ is the luminosity. One can set the truncation magnitude at $M_{\rm trunc} = -10$ corresponding to the predicted range of minimum halo-mass that can host star-forming galaxies \citep{Faucher:2011}. %However, the error bars for the best-fit function from non-parametric method are not well-defined beyond the range of training data. Therefore, we restrict ourselves to $M_{\rm trunc} = -17$. 
However, due to insufficient data at some redshifts at larger magnitudes ($M_{\rm UV}> -17$), our free-form LFs obtained using GP will be biased to the mean function beyond the range of training data. Therefore, we restrict ourselves to $M_{\rm trunc} = -17$. %Therefore, we use $M_{\rm trunc}= -17$  in our analysis.  %\da{(due to insufficient data at some redshifts for larger magnitudes).} 
Accurate estimation of $\rho_{\rm UV}$ from UV galaxies requires a careful analysis of the LF profile that can well-describe the number densities of star-forming galaxy samples down to observed limits. The best model of LF in literature is often assumed to be the Schechter function \citep{Schechter:1976},
\begin{equation}\label{eq:Schechter}
    \Phi(M) = 0.4 \,\ln10 \,\phi^{*}\,[10^{0.4(M^{*}-M)}]^{1+\alpha}\, \exp[-10^{0.4(M^{*}-M)}]
\end{equation}
parameterized by $\phi^{*}$ (Mpc$^{-3}$mag$^{-1}$), $M^{*}$ and $\alpha$. We find the posterior distributions of Schechter function parameters in redshifts $z\sim 2-12$ using the UV luminosity data sets described in~\autoref{sec:data}. We also fit same data sets using the non-parametric method discussed in~\autoref{GP}. 
We then obtain $\rho_{\rm UV}$ corresponding to each of these best-fit LFs using~\autoref{eq:Schechter}.
%\da{One can use the truncation magnitude at $M_{\rm trunc} = -10$ corresponding to the predicted range of minimum halo-mass that can host star-forming galaxies \citep{Faucher:2011}. However, the error bars for the best-fit function from non-parametric method are not well-defined beyond the range of training data. Therefore, we restrict ourselves to $M_{\rm trunc} = -17$. }

In order to constrain the reionization process, we need to adopt a parametric form \citep{Yun-Wei_Yu:2012, Ishigaki:2015, Ishigaki:2018} or a free-form \citep{Hazra:2020,Paoletti:2020ndu,Krishak:2021,Paoletti:2021gzr} of $\rho_{\rm UV}$ evolution to constrain $\dot{n}_{ion}$. The model-independent reconstruction of reionization process in \cite{Krishak:2021} rules out the single power-law \citep{Yun-Wei_Yu:2012} form  which %describes the decline of $\rho_{\rm UV}$ at low redshifts  but
that is unable to replicate the decline at $z\sim 8$. This results in an incorrect value of the Thomson scattering optical depth. Therefore, in our analysis we use the logarithmic double power law \cite{Ishigaki:2015,Ishigaki:2018} to describe $\rho_{\rm UV}$,
\begin{equation}\label{eq:log_double_power_law}
    \rho_{\rm UV} = \frac{2\rho_{{\rm UV},z=z_{\rm tilt}}}{10^{a(z-z_{\rm tilt})} + 10^{b(z-z_{\rm tilt})}},
\end{equation} 
where $\rho_{{\rm UV},z=z_{\rm tilt}}$ is the normalization factor at $z_{\rm tilt} \sim 8$, and $a$ and $b$ are the slopes. 

Once the evolution of \QHII\ from Eq.~\ref{eq:ionise_eq} is determined, we compute the reionization optical depth at redshift $z$ using
\begin{equation}\label{eq:reion_tau}
    \tau_{\rm re} = \int_{0}^{z} \frac{c(1+z^{'})^2}{H(z^{'})} \QHII(z^{'})\langle n_{H}\rangle \sigma_{T}(1 + \eta\frac{Y_p}{4X_p})dz^{'}, 
\end{equation}
where $c$ is the speed of light, $H(z)$ is the Hubble parameter, and $\sigma_T$ is the Thomson scattering cross section. Here we assume that helium is singly ionized ($\eta$ = 1) at $z > 4$ and doubly ionized ($\eta$ = 2) at $z \leq 4$ \citep{Kuhle:2012}.

\subsection{Gaussian Process Regression}\label{GP}
Gaussian process regression is a non-parametric Bayesian regression method. This method has been extensively used in cosmological data analysis~\citep{Marina_Seikel_2012,Shafieloo:PhysRevD.87.023520,Calderon:2023}. A Gaussian process is a collection of random variables such that any finite subset of these random variables has a multivariate Gaussian distribution~\citep{Rasmussen}. A GP is described by its mean and covariance functions, defined as $\mu(\mathrm{X}) = \mathbb{E}[f(\mathrm{X})]$, and
$k(\mathrm{X},\mathrm{X'}) = \mathbb{E}[(f(\mathrm{X})-\mu(\mathrm{X}))(f(\mathrm{X'})-\mu(\mathrm{X'}))]$, respectively, for a real process $f(\mathrm{X})$. In particular, here  $f(\mathrm{X})$ defines the luminosity-magnitude relation guided by data for a given mean function. The covariance function gives the covariance between two random variables and characterizes the covariance matrix having elements $C_{i,j} = k(x_{i},x_{j})$. Given a finite set of training points $\mathrm{X}=\{x_i\}$, a function $f(\mathrm{X})$ evaluated at each $x_i$ is a Gaussian random variable and the vector $\mathrm{f}(\mathrm{X})=\{f_i\}$ has a multivariate Gaussian distribution given as $\mathrm{f}(\mathrm{X})\sim \mathcal{N}\left(\mu(\mathrm{X}), C(\mathrm{X},\mathrm{X})\right)$.  
The choice of the covariance functions is important. We choose the Radial Basis Function (RBF) kernel as the covariance function for our analysis, defined as  $ K(x_i,x_j)= \sigma_{\ell} \text{exp} \left(- \frac{(x_i-x_j)^2}{2\ell^2}\right)$, where $\sigma_{\ell}$ and $\ell$ are the kernel hyperparameters. The $\sigma_{\ell}$ is the amplitude parameter that can be thought as an offset that decides the tilt of reconstructed function $f(\mathrm{X})$ from given mean function and $\ell$ describes the characteristic length of correlation. \da{The role of two hyperparameters are discussed in detail in Appendix~\ref{sec:A}}. 
Data points act as training points to optimize the hyperparameters and provide posterior prediction along with uncertainty on the predictions for the given test points. 

Although GP is a formalism to be used here to study actual trends of data to define the luminosity-magnitude relation, there are practical technicalities regarding the choice of mean function that require special care. In principle, one can choose a zero mean function or any random function to start with. In this study, however, we are particularly interested in checking the validity of the Schechter function to be a correct description of the luminosity-magnitude relation. To test that, we allow the three Schechter function parameters to vary along with the GP hyperparameters, $\ell$ and $\sigma_{\ell}$. In this way, the hyperparamter posteriors, marginalized over the Schechter function parameters can indicate, in a conservative way, whether the Schechter function is a correct model to address the observational data.

%However, often final results turn out not to be independent of this choice. The final reconstructed function especially its covariance is found to keep the memories of choice of mean function. In this study we are particularly interested to check validity of Schechter function to be correct description of luminosity-magnitude relation. Therefore, we followed two technical implementation of GP. We use best fit schechter function as a mean function and samples the GP hyperparameters. This gives us confidence level of deviation of luminosity-magnitude relation from Schchter best fit. Furthermore, in order to find possible trend driven by the data, we marginalise over all allowed Schechter function determined by its three parameters by choosing each of them to be mean function. Therefore, we reconstruct $f(\mathrm{x})$ minimising a hybrid log likelihood comprise with GP hyperparameters and three Schechter parameters. 

\section{Data sets}\label{sec:data}
We use rest frame UV luminosity function data for redshifts $z\sim 2,~3,~4,~5,~6,~7$ derived in~\cite{Bouwens:2021} using HUDF, HFF, and CANDELS fields and  HFF data compiled by \cite{Ishigaki:2018}. For $z\sim4-7$ we add data from Hyper Suprime-Cam (HSC) Subaru Strategic Program (SSP) survey \cite{Harikane22:goldrush}. These data sets are corrected for active galactic nucleus (AGN) contribution and mostly at brighter end of luminosity. For redshifts 8, 9 and 10 we use luminosity function data from~\cite{Bouwens:2021},~\cite{bowler20},~\cite{oesch18} and~\cite{McLeod16}. We also use the data redshift 9 and 12 from JWST \cite{Harikane23}.

We use neutral hydrogen fraction data to constrain reionization  from observations of Ly$\alpha$-emitting galaxies~\citep{Ono2012, Tilvi2014, Schenker2014, Mason:2018}, high-redshift quasar spectra~\citep{Schroeder:2013,Greig:2016vpu, Davies:2018pdw}, gamma ray bursts~\citep{Totani:2005ng, McQuinn2008}, dark fraction in the spectra of bright quasars~\citep{McGreer:2015} and ionized near-zones around high-redshift quasars~\citep{Bolton11,mortlock11}.
 
From Planck CMB, we use Planck binned \ensuremath{\tt Plik} TTTEEE likelihood with low multipole temperature and polarization likelihoods and the lensing likelihood as discussed in Planck baseline~\citep{Planck-VI:2020}.

\section{Results}\label{sec:results}
\subsection{Relation between magnitude and Galaxy UV LF}\label{sec:LF_form}
To characterise the galaxy UV LF, we use both Schechter function (\autoref{eq:Schechter}) and GP to fit the luminosity data. We use data for the largest possible range of magnitude, $-25<M_{\rm UV} < -14$ wherever available. We use CosmoMc~\cite{Lewis2002} as a generic sampler to explore the parameter space.

\begin{figure}
       \includegraphics[width=0.48\linewidth]{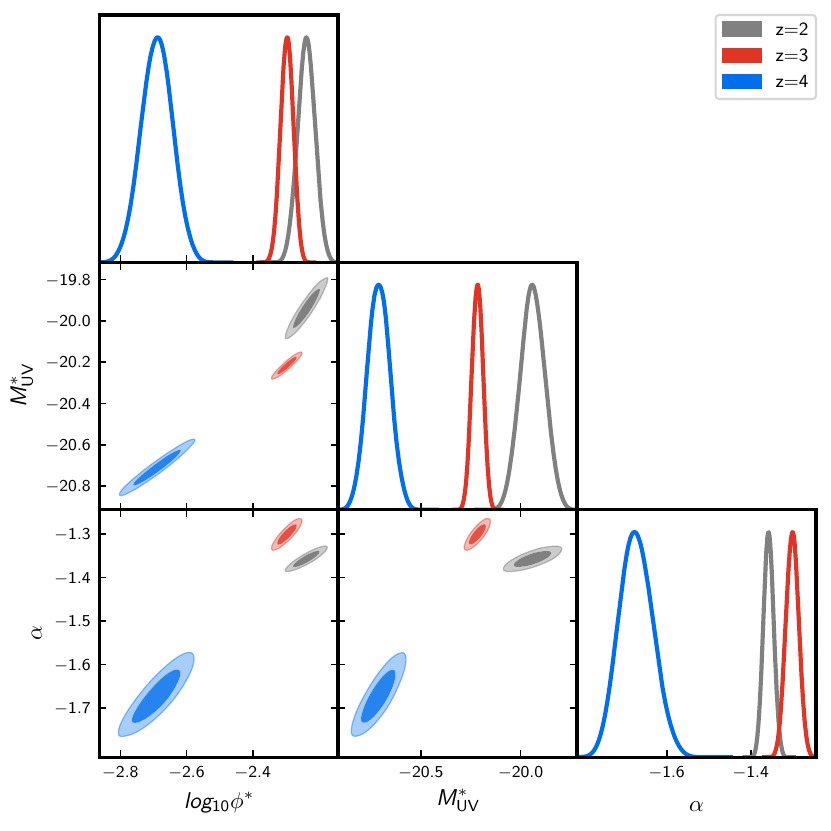}
       \includegraphics[width=0.48\linewidth]{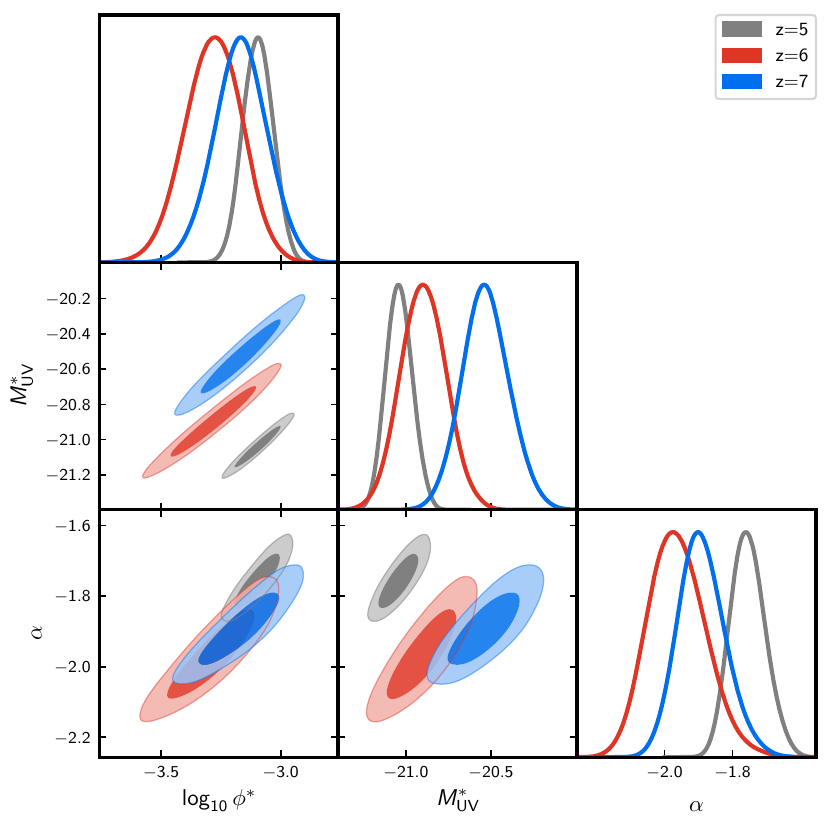}
       \includegraphics[width=0.48\linewidth]{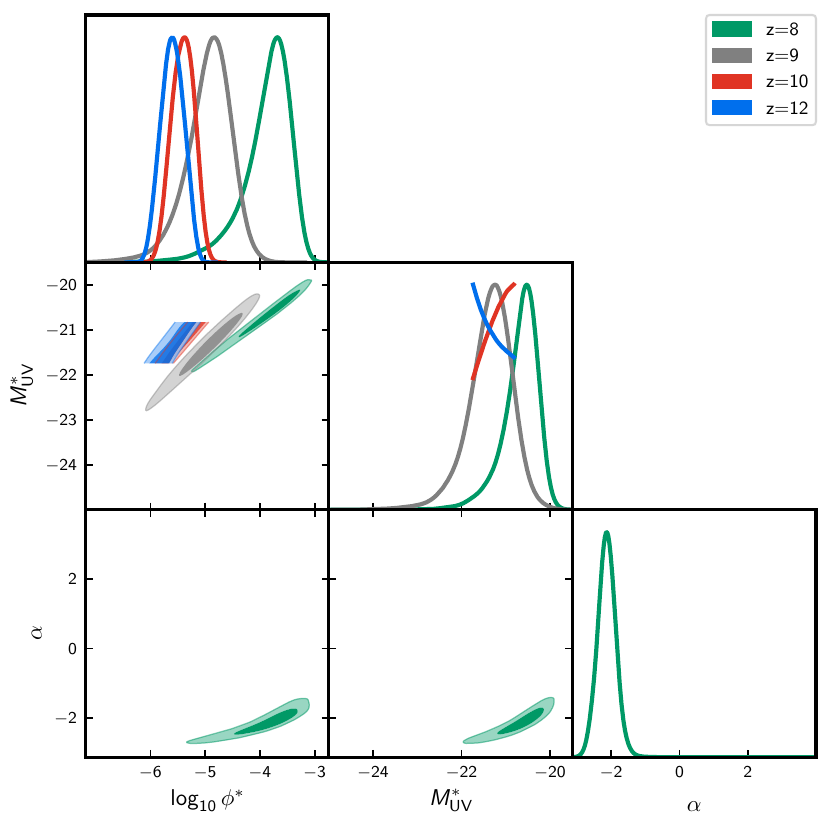}
\caption{Constraints on the Schechter function parameters using galaxy luminosity data at redshifts $z\sim 2-12$.}
\label{fig:Schechter}
\end{figure}

\begin{figure}[h!] 
\centering
\includegraphics[width=0.48\textwidth]{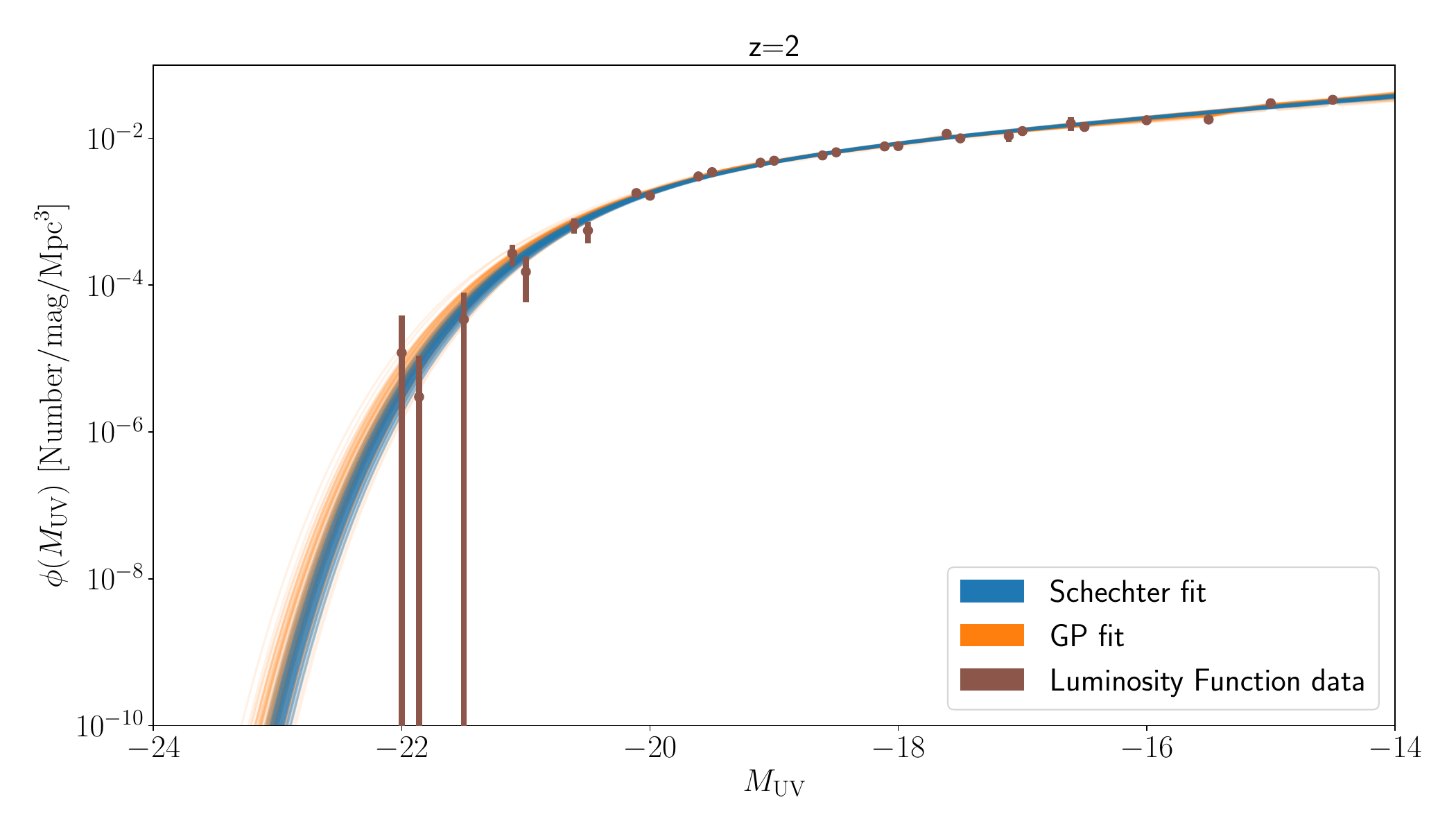}
\includegraphics[width=0.48\textwidth]{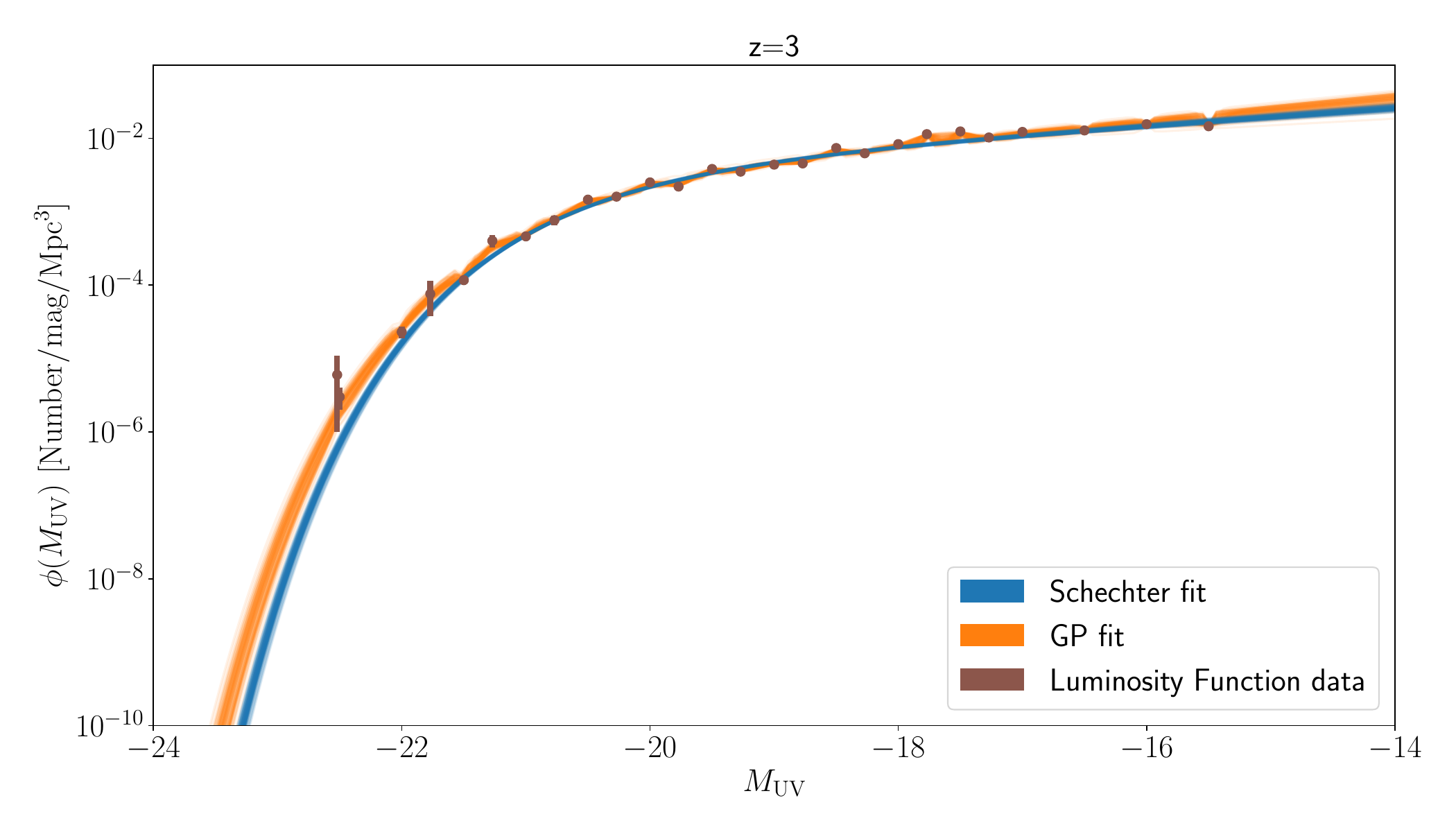}
\includegraphics[width=0.48\textwidth]{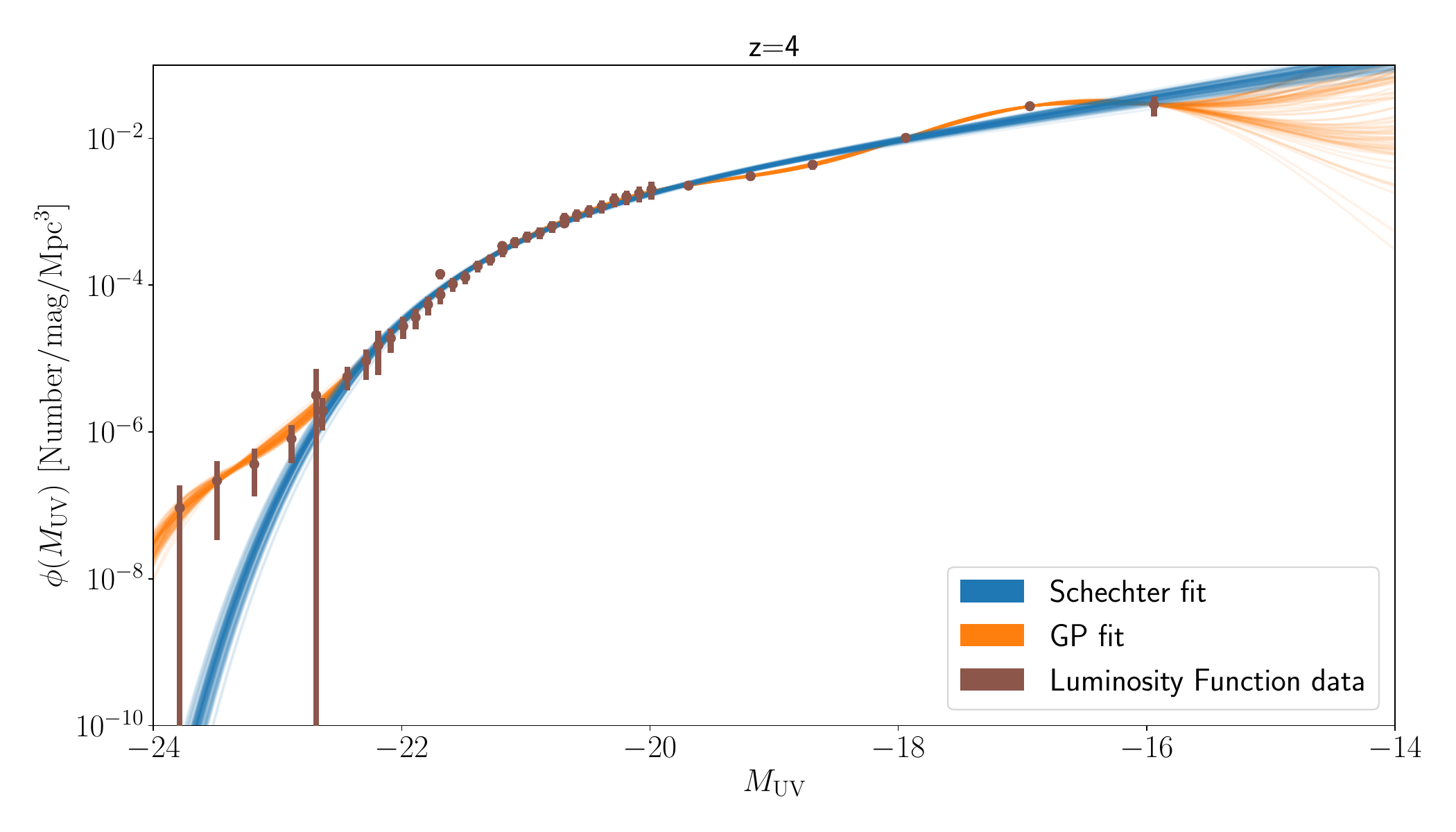}
\includegraphics[width=0.48\textwidth]{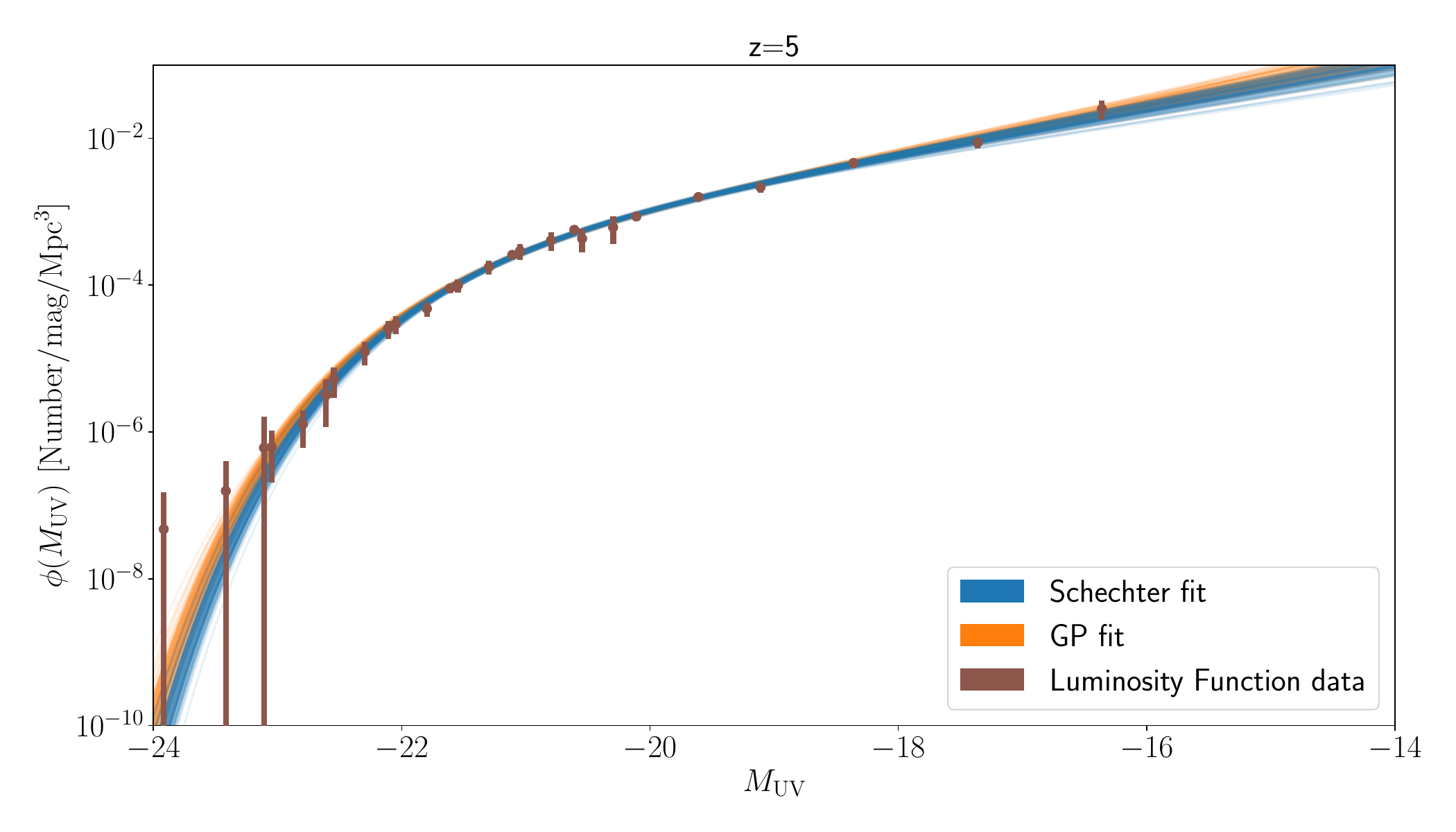}
\includegraphics[width=0.48\textwidth]{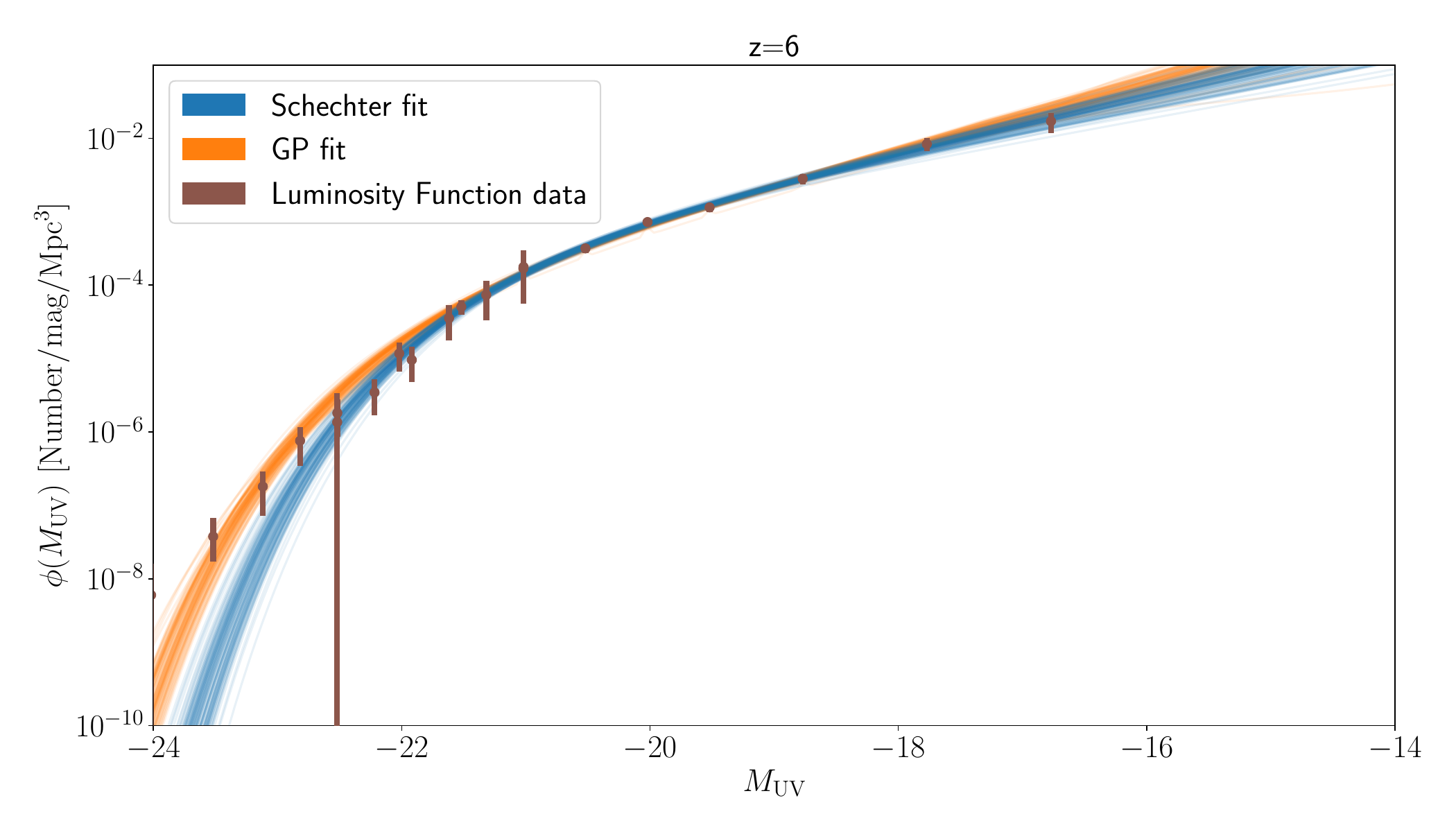}
\includegraphics[width=0.48\textwidth]{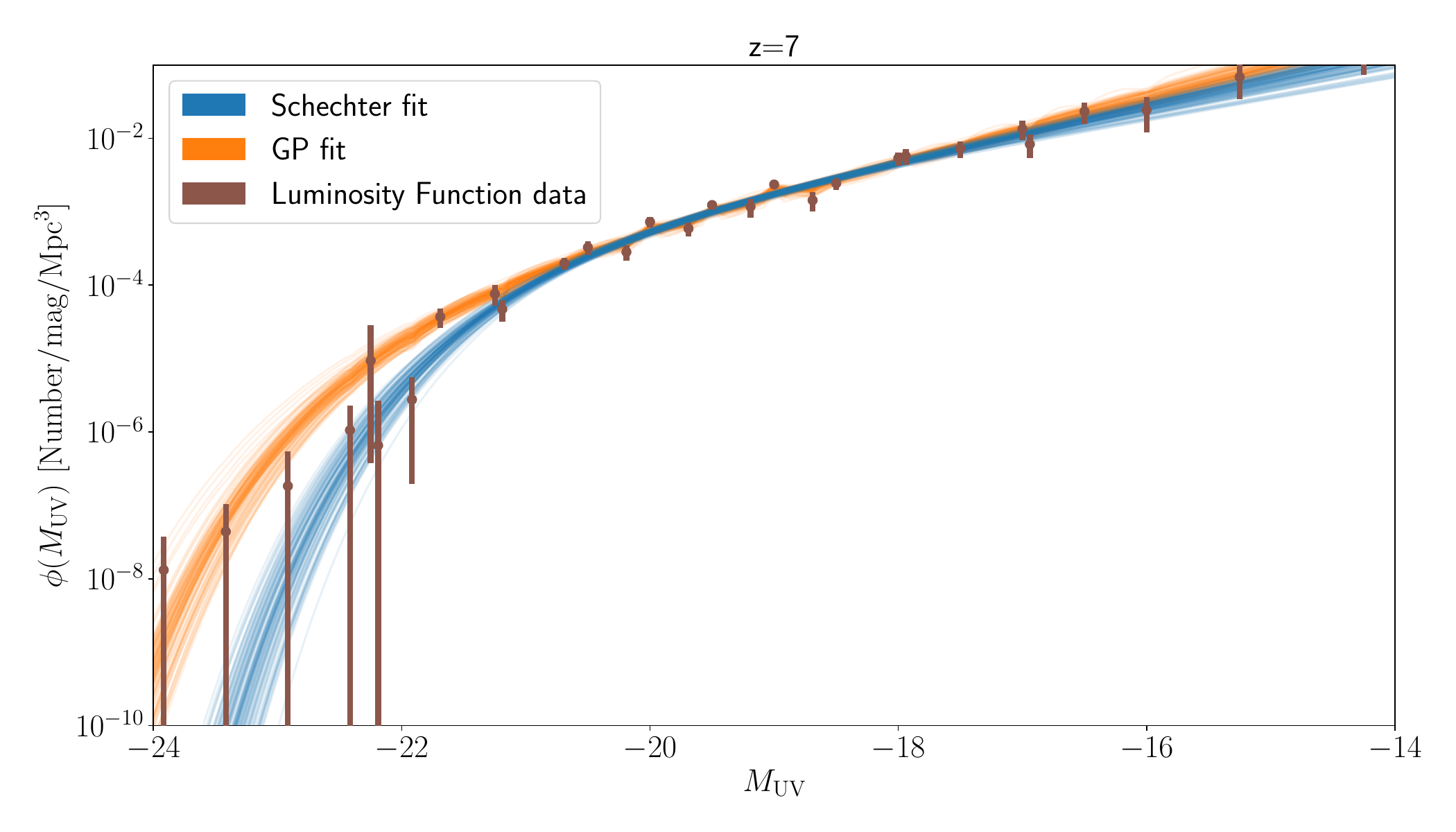}
\caption{Comparison of luminosity functions from GP and Schechter function model derived from posterior samples of corresponding parameters at redshifts $z\sim 2-7$.}
\label{fig:gp_sc_comparison1}
\end{figure}

\begin{figure}[htb!] 
\centering
\includegraphics[width=0.48\textwidth]{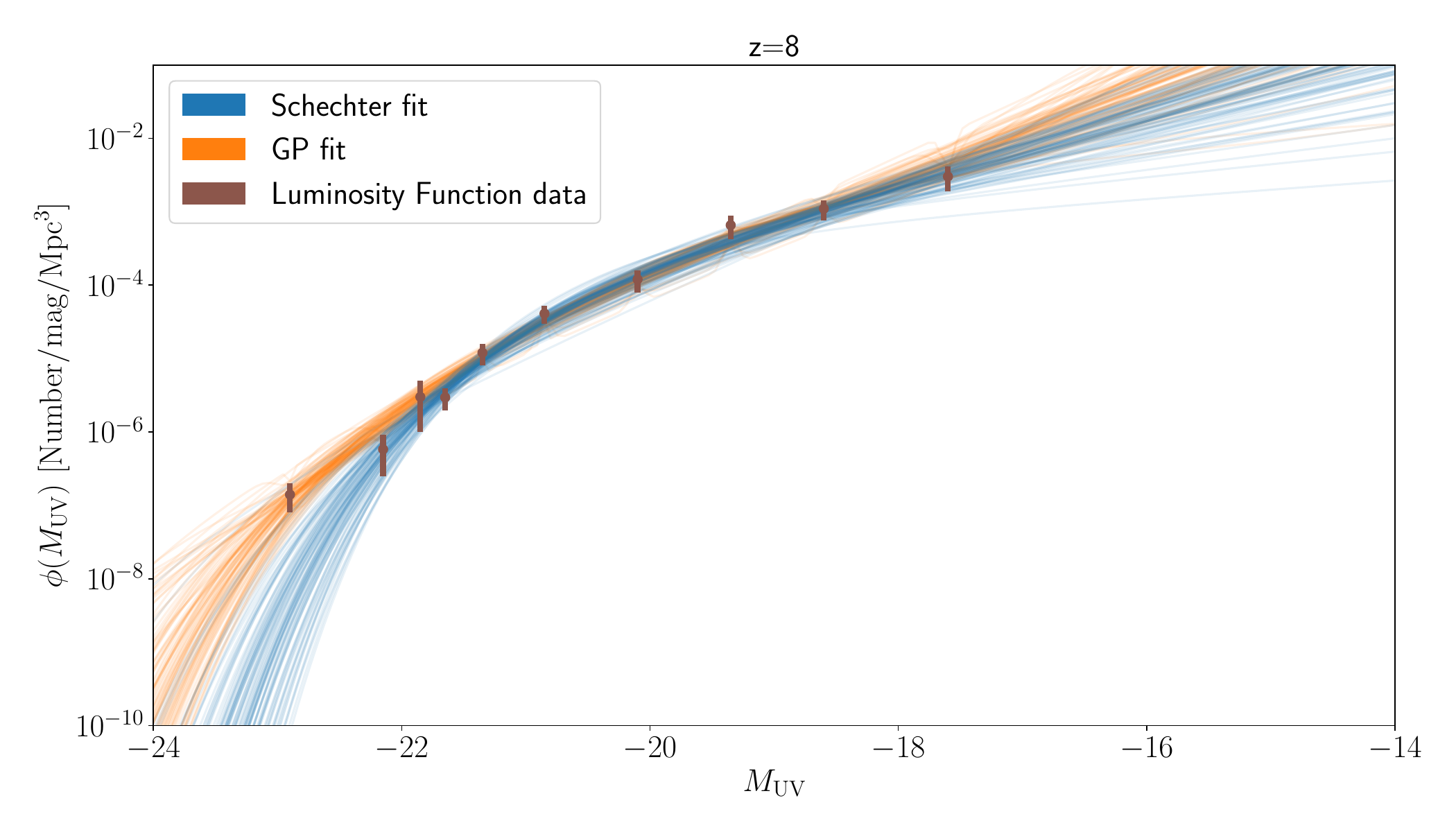}
\includegraphics[width=0.48\textwidth]{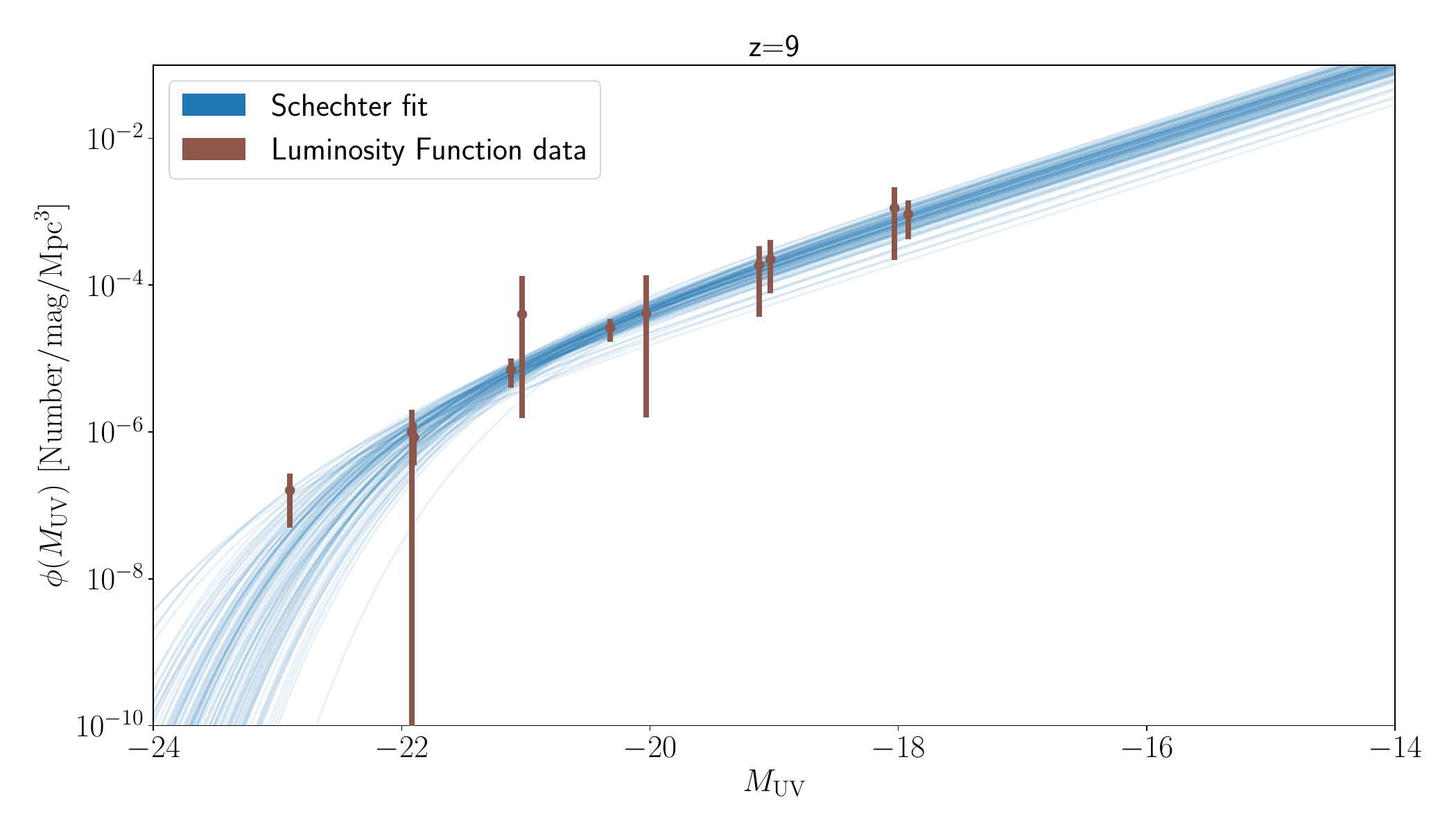}

\includegraphics[width=0.48\textwidth]{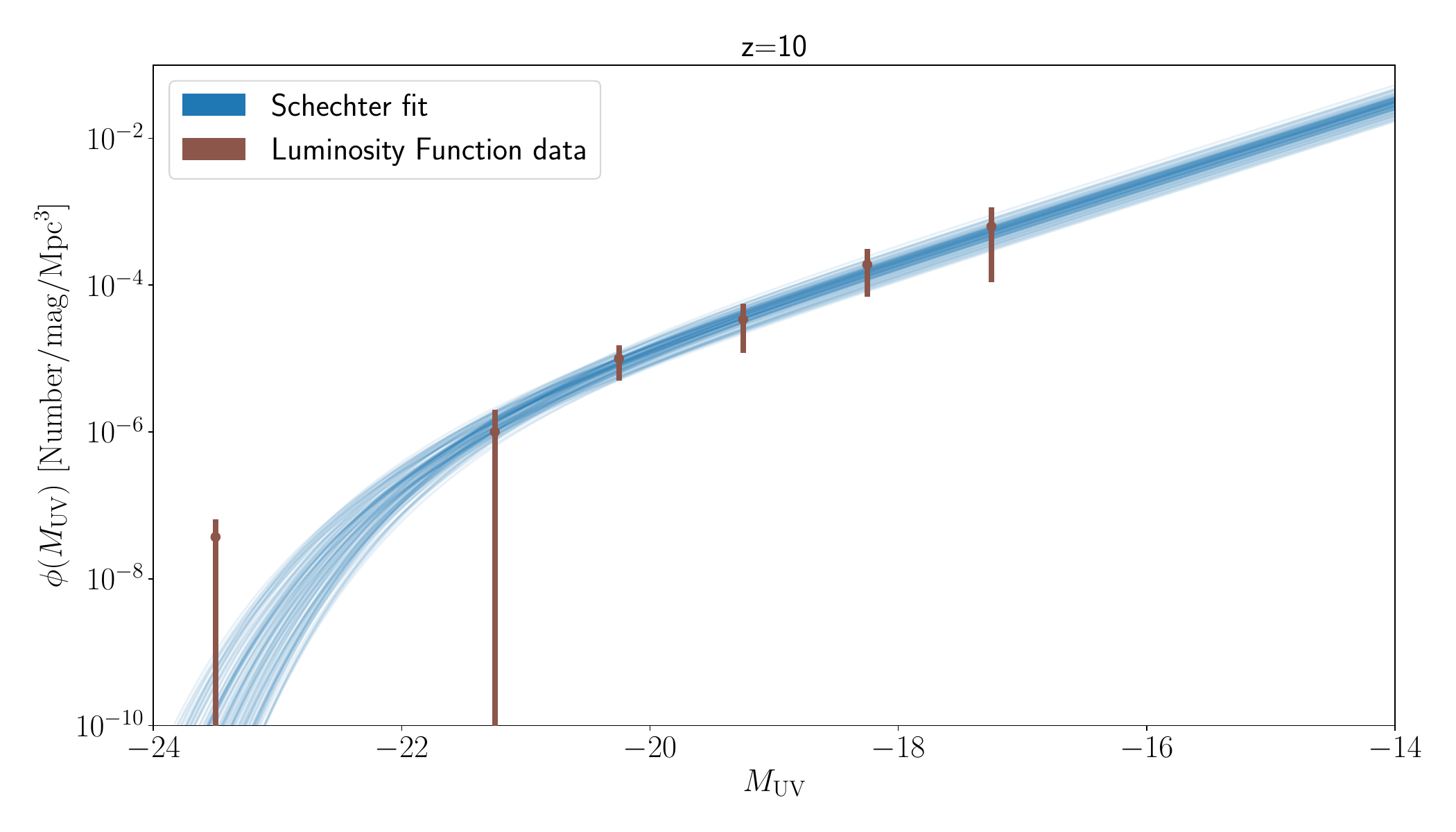}
\includegraphics[width=0.48\textwidth]{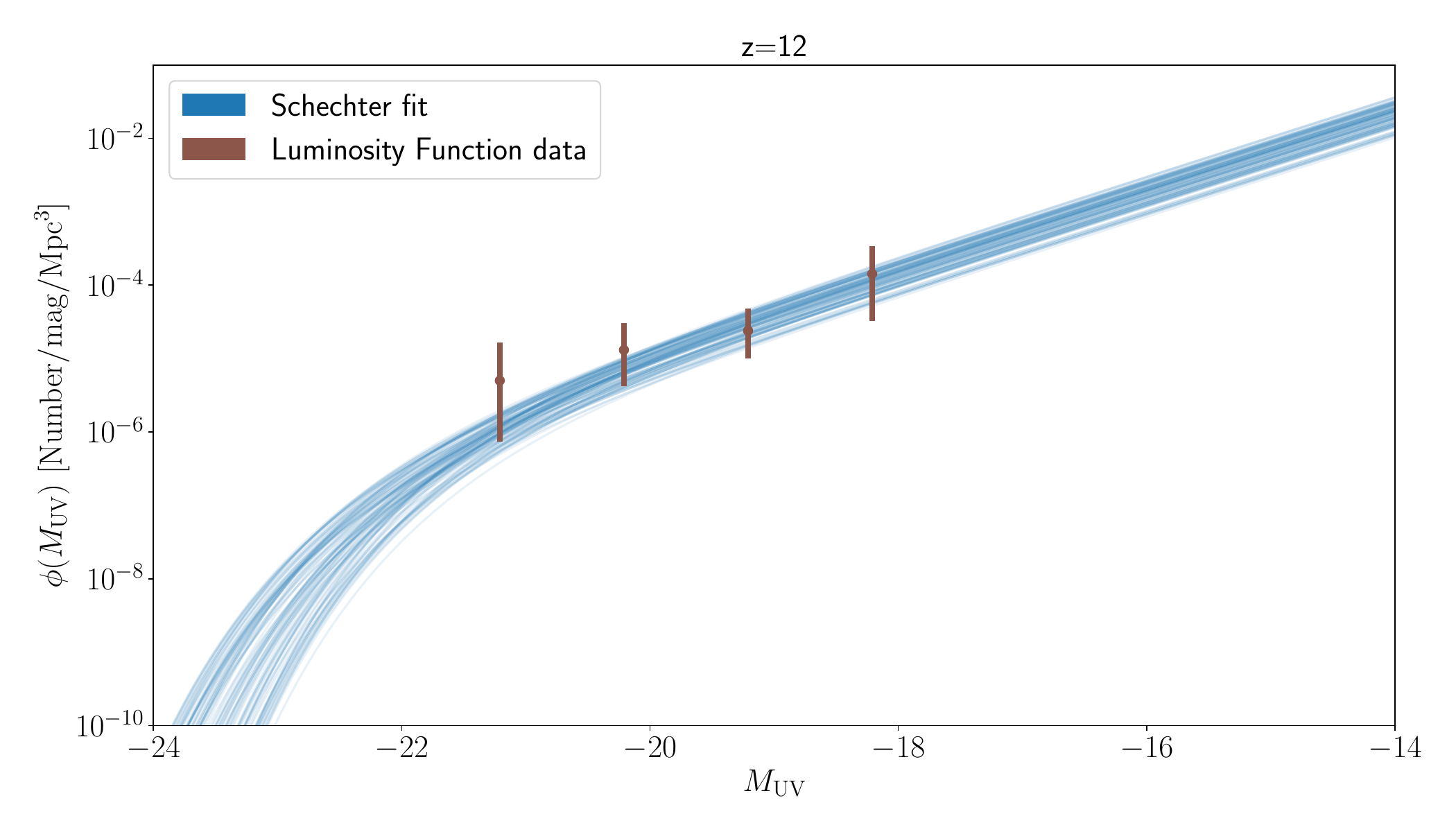}\\~\\
\caption{Same as Figure~\ref{fig:gp_sc_comparison1} for $z\sim8, 9, 10$ and $12$.}
\label{fig:gp_sc_comparison2}
\end{figure}

\begin{figure}
       \includegraphics[width=0.48\linewidth]{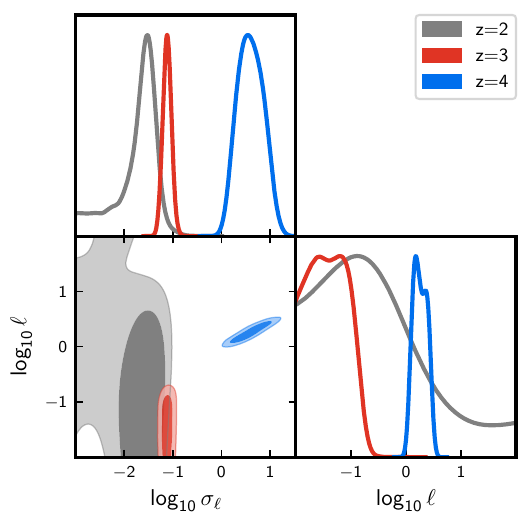}
       \includegraphics[width=0.48\linewidth]{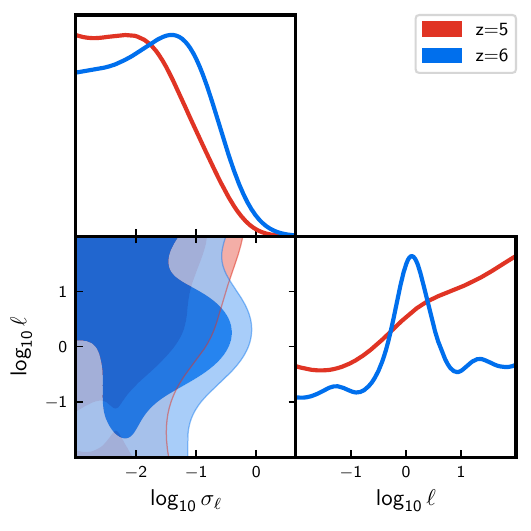}
       \includegraphics[width=0.48\linewidth]{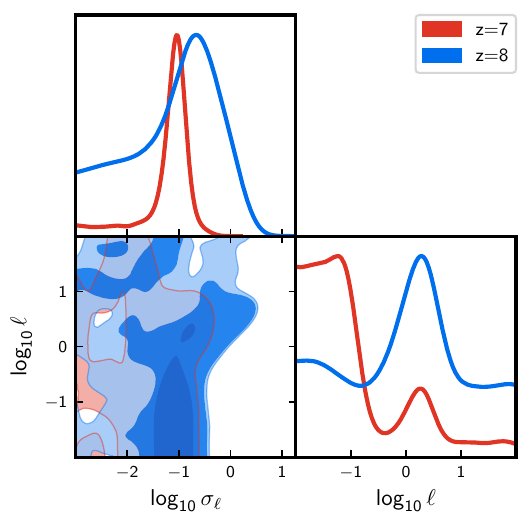}
       \includegraphics[width=0.48\linewidth]{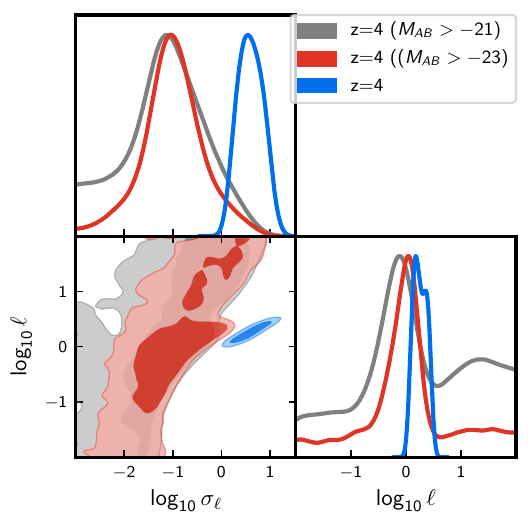}       
\caption{Constraints on the GP hyperparameters at different redshifts. Redshifts between $z=2-12$ are divided into 3 plots. At lower redshifts, mainly at $z=3$ and $4$, the data prefers significant deviation from the Schechter function. At higher redshifts, although there are some hints of deviations at $z=7,~8$, they are not statistically significant. For $z=4$, where we find most significant deviation, we reanalyze the data with two cuts $M_{\rm UV}>-23$ and $M_{\rm UV}>-21$ (bottom right plot). We notice that when we include the data from lower magnitudes, the Schechter function becomes increasingly inconsistent with the data.}
\label{fig:GP_contours}
\end{figure}

For fitting the Schechter function we keep all three parameters $M^{*}, \phi^{*}$ and $\alpha$ free. However, we notice that for $z>8$ the data is not able to constrain all the parameters. Therefore, at $z \geq 9$, we keep the slope $\alpha$ fixed at $-2.35$ following~\cite{Harikane:2022} and determine the other two Schechter parameters at $z\sim$ 9--12. We find the estimated parameters are consistent with previous results of~\cite{bowler20, Bouwens:2021, Ishigaki:2018}. For $z \sim  10$ and $12$, we fix the prior on $M^{\ast}_{UV}$ as the 68\% range obtained in $z \sim 9$ analysis. In~\autoref{fig:Schechter} we present posterior distributions of the Schechter function parameters. We fit the free-form of LF using GP with the same data sets keeping GP hyperparameters ($\ell$ and $\sigma_{\ell}$) free along with three Schechter parameters. We avoid GP fitting at $z\sim$ 9 -- 12 due to the unavailability of low magnitude high signal-to-noise data where deviation is expected and at these redshifts, even all Schechter parameters cannot be optimised.  The analysis of ten redshifts is divided into three plots.  
Random samples of luminosity function from the chains for Schechter function are plotted in blue lines in~\autoref{fig:gp_sc_comparison1} and ~\ref{fig:gp_sc_comparison2}. Modified functions from the Gaussian process are plotted in orange lines in the same figures. 
Deviations from Schechter function below magnitude $M_{\rm UV}=-21$ can be noticed. A similar deviation is also reported in~\citep{Ono:2018}. It is important to note that most of the previous studies are based on HFF and HUDF data~\citep{Finkelstein:2015,Bouwens:2015,Ishigaki:2018, Bouwens:2021} and others ground-based observations \citep{bowler20} that are limited by the magnitude larger than $-23$. According to those studies, Schechter function is the best description of the magnitude-luminosity relation. In our study, we combine those previous data sets with ancillary luminosity data sets of brighter galaxies ($M_{\rm UV} \lesssim -23 $) from HSC data \citep{Harikane:2022}. Above this magnitude, the AGN contamination to the galaxy luminosity is negligible~\citep{Glikman:2011,Giallongo:2015,Niida:2016,Akiyama:2017, McLure:2017} and therefore the previous results are consistent with our results up to certain magnitudes. The added HSC data of brighter galaxies are corrected for contribution from AGNs~\citep{Harikane:2022}. %The random samples presented in Figure~\ref{fig:gp_sc_comparison1} and ~\ref{fig:gp_sc_comparison2} show deviations at the brightest end of the luminosity at all redshifts. 
%To display the deviations more clearly, in~\autoref{fig:residual} we plot the residuals {\it w.r.t.} the best fit Schechter function. 
We find maximum excess in the bright end shape of galaxy UV LF instead of expected exponential drop of Schechter function at all redshifts where low magnitude ($M_{\rm UV} \lesssim -21$) data are available.  In certain redshifts, due to high signal-to-noise ratio, the deviation is detected at high statistical significance while for other redshifts we obtain a similar deviation with lower significance. Importantly, at $z=3,4$ the hyperparameters posteriors plotted in the upper left panel of~\autoref{fig:GP_contours} indicate that the Schechter function (used as a mean function in this analysis) is ruled out by the data at high significance. Data from $z=2$ and $7$ also prefer modification over Schechter function at around $\sim$95\% confidence level (C.L). In order to understand the source for this deviation, we reanalyse $z=4$ data with two different data cuts (with $M_{\rm UV}>-23$ and $M_{\rm UV}>-21$). The results are presented in the lower right panel of~\autoref{fig:GP_contours}. When the brightest part of the observation $-24<M_{\rm UV}<-23$ is not used, we notice a decrease in the significance in the GP hyperparameters ruling out Schechter function at 3$\sigma$. When we use more conservative cuts by masking data between $-24<M_{\rm UV}<-21$, we find that further drop in significance. These tests reveal that the modifications to Schechter function are needed by the luminosity observations at the lowest magnitude (the brightest) objects.

  %The maximum (minimum) significance values of deviation from Schechter best fit at brightest end of LF ($M_{\rm UV}\lesssim -23$) are 2.07$\sigma$ (1.04$\sigma$), 1.47$\sigma$ (0.05$\sigma$), 0.99$\sigma$ (0.01$\sigma$), 1.81$\sigma$ (0.05$\sigma$), 0.99$\sigma$ (0.2$\sigma$), and 1.0$\sigma$ (0.98$\sigma$) at z $\sim$ 3, 4, 5, 6, 7 and 10.

%##########

\begin{table}[!htb]
\begin{center} 
\begin{tabular}{|c|cccc|c|}
\hline
Model & \multicolumn{4}{c|}{Schechter fit}                                                                                                               & \multirow{2}{*}{\begin{tabular}[c]{@{}c@{}}Gaussian process\\ $\log_{10}\rho_{\rm UV}$\end{tabular}} \\ \cline{1-5}
$z$     & \multicolumn{1}{c|}{$\log_{10}\phi^{\ast}$} & \multicolumn{1}{c|}{$M^{\ast}_{\rm UV}$} & \multicolumn{1}{c|}{$\alpha$} & $\log_{10}\rho_{\rm UV}$ &                                                                                                      \\ \hline
2     &\multicolumn{1}{c|}{$-2.238\pm 0.026$ }                      & \multicolumn{1}{c|}{$-19.940\pm 0.061$}                    & \multicolumn{1}{c|}{ $-1.357\pm 0.012$}         &                       $26.426\pm0.010$   & $26.436\pm 0.012$                                                                                                      \\ \hline
3     & \multicolumn{1}{c|}{ $-2.297\pm 0.019$}                      & \multicolumn{1}{c|}{$-20.216\pm 0.027$}                    & \multicolumn{1}{c|}{$-1.301\pm 0.015$}         &                          $26.4741\pm 0.0063$&   $26.489^{+0.014}_{-0.012}$                                                                                                   \\ \hline
4     & \multicolumn{1}{c|}{$-2.689\pm 0.047$}                      & \multicolumn{1}{c|}{$-20.711\pm 0.057$}                    & \multicolumn{1}{c|}{$-1.673\pm 0.040$}         &$ 26.470\pm 0.014$                          &  $26.470\pm0.013$                                                                                               \\ \hline
5     & \multicolumn{1}{c|}{$-3.095\pm 0.062$}                      & \multicolumn{1}{c|}{$ -21.039\pm 0.077$}                    & \multicolumn{1}{c|}{$-1.755^{+0.047}_{-0.055}$}         &$ 26.264\pm 0.019$                          & $26.275\pm 0.016$                                                                                                \\ \hline
6     & \multicolumn{1}{c|}{$-3.28\pm 0.12$}                      & \multicolumn{1}{c|}{$ -20.90\pm 0.13$}                    & \multicolumn{1}{c|}{$-1.962^{+0.076}_{-0.091}$}         &$26.165^{+0.033}_{-0.028}$                          &     $26.188^{+0.022}_{-0.026}$                                                                                            \\ \hline
7     & \multicolumn{1}{c|}{$-3.17\pm 0.11$}                      & \multicolumn{1}{c|}{$ -20.53\pm 0.14$}                    & \multicolumn{1}{c|}{$-1.890^{+0.062}_{-0.074}$}         &$26.039\pm 0.022$                          &  $26.071^{+0.034}_{-0.023}$                                                                                           \\ \hline
8     & \multicolumn{1}{c|}{$-3.88^{+0.51}_{-0.23}$}                      & \multicolumn{1}{c|}{$-20.68^{+0.46}_{-0.24}$}                    & \multicolumn{1}{c|}{$-2.13\pm 0.27$}         &$25.616^{+0.099}_{-0.066}$                          &  $25.705\pm 0.084$                                                                                           \\ \hline
9     & \multicolumn{1}{c|}{$-4.94^{+0.46}_{-0.31}$}                      & \multicolumn{1}{c|}{$-21.36^{+0.54}_{-0.38}$}                    & \multicolumn{1}{c|}{$-2.35$}         &$25.12^{+0.16}_{-0.11}$                          &  --                                                                                           \\ \hline
10     & \multicolumn{1}{c|}{$-5.79^{+1.1}_{-0.84}$}                      & \multicolumn{1}{c|}{$-21.8^{+1.6}_{-1.2}$}                    & \multicolumn{1}{c|}{$-2.35$}         &$24.51^{+0.19}_{-0.15}$                          &  --                                                                                           \\ \hline
12     & \multicolumn{1}{c|}{$-6.19^{+0.67}_{-1.3}$}                      & \multicolumn{1}{c|}{$< -19.1$}                    & \multicolumn{1}{c|}{$-2.35$}         &$24.40^{+0.23}_{-0.12}$                          &  --                                                                                           \\ \hline

\end{tabular}
 \caption{Constraints on luminosity density obtained from Schechter fit and the GP fit to the data at different redshifts between $2<z<8$. For $z=9,~10,~12$ we just provide only the Schechter fit as even the three parameters can not be constrained with two tailed distribution from the data, due to low signal-to-noise ratio.} \label{table:gp_rho_uv}
\end{center} 
\end{table}
%$$$$$$$$

%$$$$$$
\begin{figure}
\includegraphics[width=\linewidth]{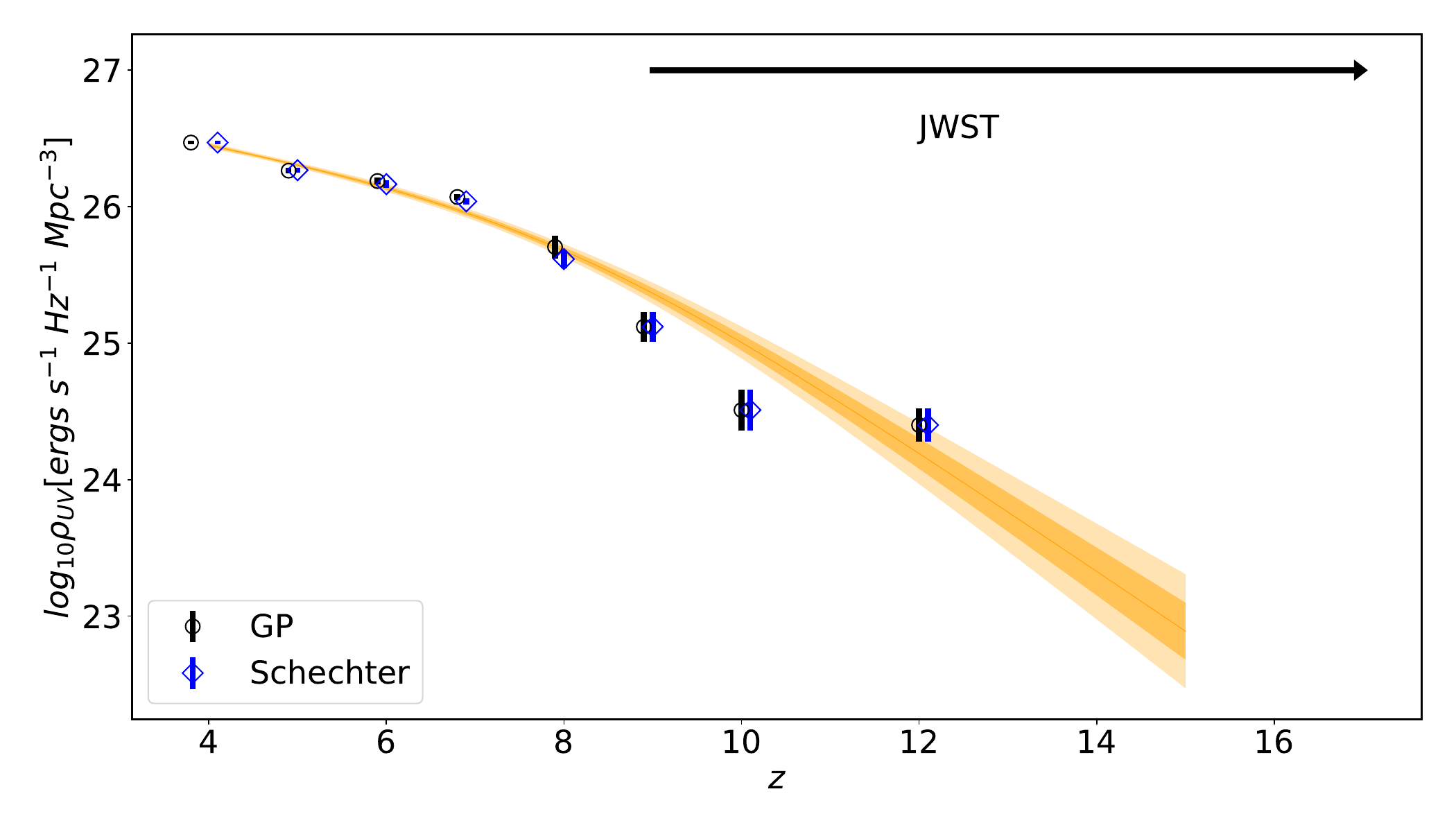}
%\includegraphics[width=0.48\linewidth]{figs/tau_crop.pdf}
%\includegraphics[width=0.48\linewidth]{figs/tau_dist_hybrid.pdf}
%\centering
%\begin{multicols}{3}
%       \includegraphics[width=\linewidth]{figs/rho_uv_samples_dpl_params_Mx17_v1.pdf}\par
 %      \includegraphics[width=\linewidth]{figs/QHII_fginv_sc_gp_HYBRID.pdf}\par
 %      \includegraphics[width=0.95\linewidth,height=0.72\linewidth]{figs/tau_dist_hybrid.pdf}
%\end{multicols}
\caption{Luminosity density evolution across redshift $z\sim4-12$. Black circles and blue diamonds are $\rho_{\rm UV}(z)$s obtained from integrating the luminosity functions from GP and Schechter respectively down to the magnitude $M_{\rm UV}$ = --17. The orange line is best-fit logarithmic double power-law. The shaded regions are 1$\sigma$ and 2$\sigma$ confidence level to the logarithmic double power-law fit to $\rho_{\rm UV}$ obtained from Schechter function fit.}%. Right panel: The distributions of reionization optical depth derived for two $\rho_{\rm UV}$ data sets.}
\label{fig:dpl_tau_qhii}
\end{figure}

\begin{figure}
\includegraphics[width=\linewidth]{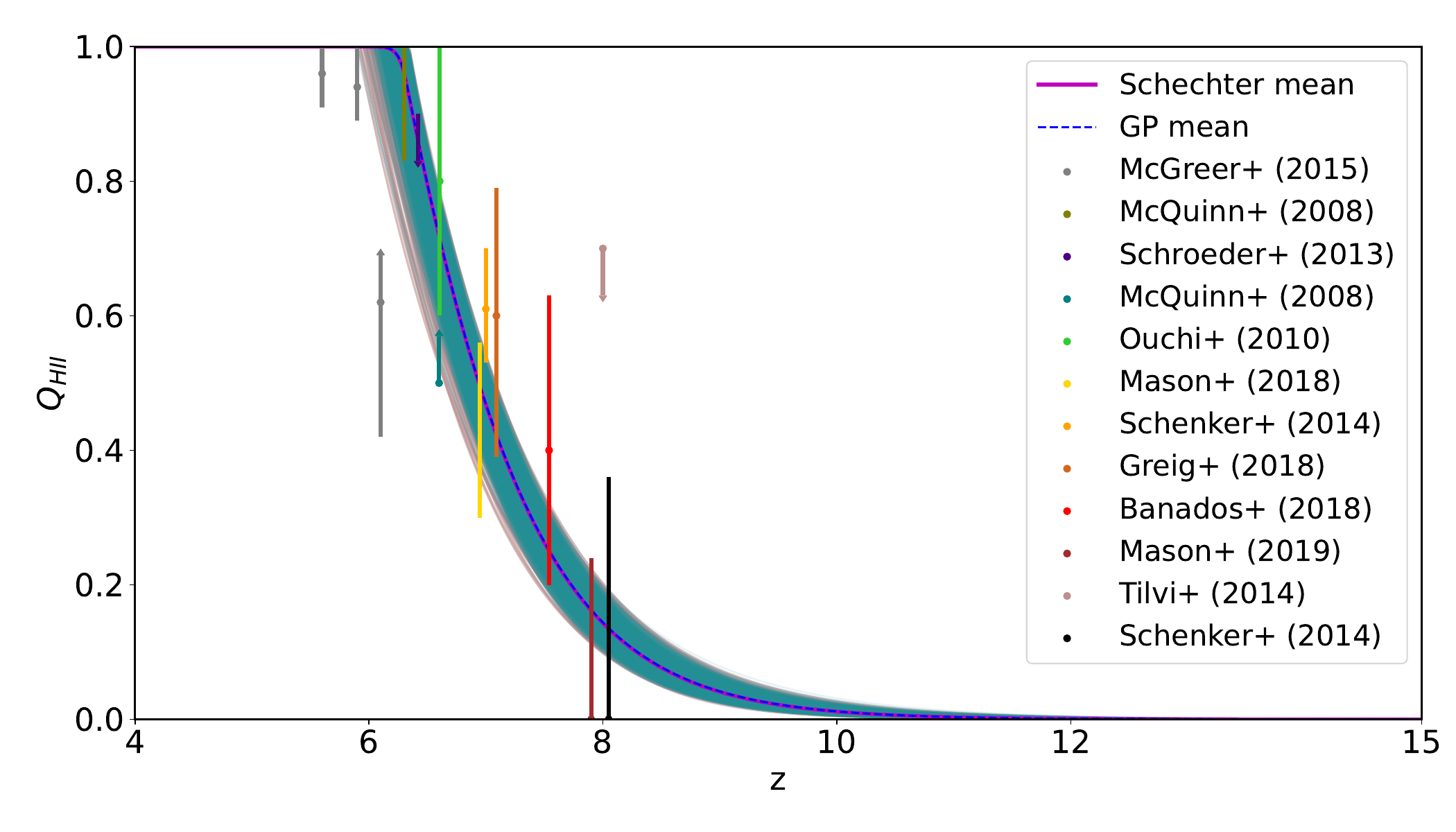}
%\centering
%\begin{multicols}{3}
%       \includegraphics[width=\linewidth]{figs/rho_uv_samples_dpl_params_Mx17_v1.pdf}\par
 %      \includegraphics[width=\linewidth]{figs/QHII_fginv_sc_gp_HYBRID.pdf}\par
 %      \includegraphics[width=0.95\linewidth,height=0.72\linewidth]{figs/tau_dist_hybrid.pdf}
%\end{multicols}
\caption{Redshift evolution of \QHII\ across $z\sim$4 --15. The dashed blue and solid magenta lines are best-fit \QHII\ estimated using $\rho_{\rm UV}$s from GP and Schechter functions respectively. The gray (cyan) curves are random samples of \QHII\ for these two respective cases. The \QHII\ data sets used for joint fitting are also shown.}
\label{fig:qhii}
\end{figure}

\subsection{Evolution of UV Luminosity Density}\label{sec:rho_uv}
We compute the UV luminosity density $\rho_{\rm UV}$ following~\autoref{Eq:l_density} integrating down to $M_{\rm UV} = -17$. The samples of LFs from  Schechter and GP analyses are used to estimate the posterior distribution of the derived parameter $\rho_{\rm UV}$. The mean and the 68\% bounds are provided in~\autoref{table:gp_rho_uv}. The changes in luminosity density compared to the Schechter function model are not noticeable in the GP results. Though Schechter function is ruled out by the data at certain redshifts, the major required modifications are noticed at the brightest end of the function and the brightest end does not contribute significantly to the integral of the UV luminosity density as the function drops logarithmically with the increase in brightness. In~\autoref{fig:dpl_tau_qhii}, we plot the redshift evolution of the luminosity density obtained from the GP (and Schechter fit) and a logarithmic double power law fit to the data (for $z \geq 6$). We demonstrate only 1$\sigma$ and 2$\sigma$ spread for logarithmic double power-law fit to luminosity density obtained from the Schechter fit. We notice outliers in the data {\it w.r.t.} the model around redshifts 9 and 10. Recent JWST data suggests an enhanced population of star-forming Galaxies above redshifts $z\sim9$  \cite{Finkelstein:2023, Eisenstein:2023}. Compared to~\cite{Krishak:2021} new data may seem to hint towards a modification to the luminosity density evolution model. However, note that the data from JWST can not constrain all of the Schechter function parameters due to low signal-to-noise ratio detection, in particular, the slope remains unconstrained (see,~\autoref{fig:Schechter}). This may bias the estimation of the luminosity density and therefore, with this data we avoid the exploration for the possible deviation from the double power law model. We expect to revisit this issue in the future when new data from JWST are made available.

%Hereafter, we nomenclate these derived UV luminosity densities $\rho_{UV,Sc}$ and $\rho_{UV, GP}$ respectively. The estimated values are quoted in Table~\ref{table:gp_rho_uv}. 
%In Figure~\ref{fig:LF_SC_GP_p1}, we find the Schechter function prescribes lower galaxy number density at bright-end of the magnitude as compared to GP. However,  due to small discrepancies between prescription of two methods,  the same is not reflected in $\rho_{\rm UV}$ values. We plot the both $\rho_{\rm UV}$ along with 1$\sigma$ errorbars in left panel of Figure~\ref{fig:dpl_tau_qhii} for comparison. %The $\rho_{\rm UV}$ values are little higher mainly visible at high redshifts when computed from GP as compared to same from Schechter function. This is because due to lack of brighter galaxy data, both Schechter and GP best fit seems same at lower redshifts. 
%We check whether we can approximate the redshift evolution of $\rho_{\rm UV}$ with double power-law in Equation~\ref{eq:log_double_power_law}. We find the minimised $\chi^2$ per degrees of freedom is 1.19 and data points are consistent with best fit function within 2$\sigma$ as seen in left panel of Figure~\ref{fig:dpl_tau_qhii}. The solid blue line represent the best fit double power-law function and the shaded regions are 1$\sigma$ and 2$\sigma$ confidence regions of posterior distribution of $\rho_{\rm UV}$.  We find the $z_{\rm tilt} = 8.54 \pm 0.18$ for $\rho_{UV, GP}$ which is shifted on higher side as compared to same for $\rho_{UV, Sc}$ which is $8.11\pm 0.51$.

%$$$$$$$$

%$$$$$$
\subsection{Constraints on Reionization}\label{sec:reion}
We obtain the constraint on redshift evolution of IGM neutral hydrogen fraction from joint fitting of Planck data, \QHII\ data, and UV luminosity density data derived in ~\autoref{sec:rho_uv}. We treat the four parameters in logarithmic double power-law form of $\rho_{UV}$ and $\log_{10}\langle f_{esc}\xi_{ion}\rangle$ as free parameters. For $C_{\rm HII}$ we use a fixed value of 3. 

In~\autoref{fig:qhii} we compare the \QHII\ evolution across redshift $z\sim4-15$ solving~\autoref{eq:ionise_eq} using $\rho_{\rm UV}$ data sets obtained from both parametric and non-parametric methods. The blue dashed line and magenta solid line are the best fit \QHII\ obtained with $\rho_{\rm UV}$ from GP and Schechter fits respectively. The grey and cyan lines indicate random samples of \QHII\ obtained from these fits respectively. We also overplot the \QHII\ data set used in our analysis. Both the $\rho_{\rm UV}$ data sets provide similar constraints on reionization history. This is not surprising as we discussed that the obtained luminosity density follows similar redshift evolution owing to the similar nature of luminosity functions at the faint source end. This result suggests that the reionization process is mainly driven by fainter galaxies. The excess dropout galaxies at the brighter-end of magnitude have limited contribution to the reionization process due to lower number density or possibly low escape fraction. Earlier publications in~\cite{Mitra:2015,Mitra:2018,Finkelstein:2019, dayal:2020, Atek:2024} reached to a similar conclusions using model-based reionization study. As shown in~\autoref{fig:qhii} we find much tighter bound of \QHII\ as compared to previous works in~\cite{Mitra:2015, Mitra:2018, Hazra:2020,Krishak:2021, Mitra:2023}. We quote the midpoint (50\%) redshift of reionization $z_{\rm re}$ in~\autoref{table:tau_zre_deltz} which is less than the Planck reported value \citep{Planck-VI:2020}. Our constraints on the duration of reionization (redshift difference between 10\% to 90\% reionization) are $\Delta z\sim 1.627^{+0.059}_{-0.071}$ and $1.627^{+0.060}_{-0.070}$ with 68\% confidence interval using Schechter function and GP based $\rho_{\rm UV}$ data sets respectively. This is consistent with a upper bound of $\Delta$z$<$2.8 as reported in \cite{Planck-xlvii:2016} from the Kinetic Sunyaev–Zel’dovich effect. This implies our values suggest a sharper reionization history. Finally our constrain on reionization optical depth is $\tau_{\rm re} = 0.0492^{+0.0008}_{-0.0006}$ and $0.0494^{+0.0007}_{-0.0006}$ from Schechter and GP analyses. These values are also listed in Table~\ref{table:tau_zre_deltz}. However, we would like to highlight here that the uncertainties in the reionization histories quoted in this subsection are underestimated as we have fixed both the clumping factor and the escape fraction in our analyses. Therefore, these summary statistics on optical depth and the duration of reionization should be used with caution in any subsequent analysis. The exercise in this subsection has been performed to indicate that the constraints on the cosmology remain unaltered even though Schechter function is ruled out by the observational data. The flattening/increase in the luminosity density data at high reshifts as indicated by the JWST data in this analysis does not show an impact in the optical depth constraints as -- (1) we have not used the very high redshift observations ($z\sim 14,~16$) from JWST, and (2) the double power-law form used for the luminosity density does not capture its increasing trend at high redshifts.
%$$$$$$$$$$
\begin{table}
\begin{center}
    
\begin{tabular}{|c|c|c|}
\hline
 & using $\log_{10}[\rho_{\rm UV}]$ from Schechter &using $\log_{10}[\rho_{\rm UV}]$ from GP\\
\hline
$\tau_{\rm re}$ & 0.0492$^{+0.0008}_{-0.0006}$&0.0494$^{+0.0007}_{-0.0006}$\\
$z_{\rm re}$&$6.778^{+0.054}_{-0.029}$ & $6.777^{+0.055}_{-0.028}$\\
$\Delta z$  & $1.627^{+0.059}_{-0.071}$& $1.627^{+0.060}_{-0.070}$\\
\hline 
\end{tabular}
 \caption{Summary of constraints on reionization history using $\rho_{\rm UV}$ data sets obtained from both parametric and non-parametric methods.} 
\label{table:tau_zre_deltz}
\end{center}

\end{table}

\section{Summary}\label{sec:summary}
In this paper, we investigate the UV luminosity functions between redshifts $z\sim2-12$ over a wide range of magnitude $-25\lesssim M_{\rm UV} \lesssim -14$ based on HST, HSC and JWST data sets. We test whether the conventional Schechter function is a valid theoretical model that can address the \da{trends of luminosity data} for such a wide range of magnitude and redshifts; and what are the implications of any modified function to the reionization history.

We fit both commonly used Schechter function model at all redshifts $z\sim2-12$ and perform a free-form reconstruction at redshifts $z\sim2-8$ using Gaussian process to the same data sets. We find although Schechter function is a very good description of the dropout galaxies for the fainter end of magnitude ($M_{\rm UV} \gtrsim -21$), its exponential tail is inconsistent with brighter dropout galaxies at almost all redshifts where low magnitude data are available. The Gaussian process regression allows a free-form reconstruction of UV LFs and therefore can well describe the excess LF at the bright-end which is \da{not addressed by Schechter function.} \da{The GP hyperparameters confirms the rejection of Schechter function at high significance at redshifts $z= 3$ and 4. Analysis with the data below magnitude of $-23$ shows the significance is more than 3$\sigma$ for $z=4$. The luminosity data at $z=2$ and 7 also disfavors the Schechter function at $\sim$ 95\% C.L. A hints of deviation from Schechter function for other redshifts is addressed with less significance due to unavailability of high quality data at low magnitudes. While these results are consistent with the deviation addressed in model dependent approaches of \cite{Ono:2018, Stevans:2018, bowler20, Harikane:2022} the GP LFs have also addressed the significant transient deviations from Schechter function $z = 3$ and 4 (Figure~\ref{fig:gp_sc_comparison1}).}  We obtain the UV luminosity densities integrating over both LF forms down to the magnitude $M_{\rm UV}  = -17$. Since at the fainter end of the magnitude, Schechter function is consistent with the data and the fainter end contributes dominantly to the luminosity density integral due to the maximum availability of Galaxies, we find similar luminosity densities in both the methods.
Therefore, reionization history is found to be similar in both cases. This implies that brighter dropout galaxies have insignificant contributions in reionization process (supported by earlier publications as well). However, integrating to more fainter sources $M_{\rm UV} > -15$ can indicate certain differences which we do not explore by considering $M_{\rm UV} = -17$ as the conservative choice. 

The data at redshifts 9, 10 and 12 do not have a high signal-to-noise ratio and therefore can not constrain all the Schechter function parameters. Therefore testing a modification to Schechter function is beyond the scope of this paper with the available data. More observational data from JWST NIRSpec will be required for further studies at higher redshifts.  Furthermore, we have not incorporated redshift dependencies of clumping factor and escape fraction into our simplistic reionization model and also neglect the contributions from quasars as reionization sources which could potentially enhance the accuracy of our findings. It would be intriguing to incorporate all these modifications into our model and reassess the current analysis when new data from JWST and other sources with significant improvement in the signal-to-noise ratio are made available. We defer these extensions for future investigations.

%Using the standard analytical reionization model, we calculate the physical parameters related to reionization and logarithmic double power-law jointly fitting derived UV luminosity densities, neutral hydrogen fraction data from galaxy, quasar, and gamma-ray burst observations and optical depth from Planck observation.
%We find little longer duration of reionization and larger reionization optical depth while using free-form UV luminosity density data as compared to same derived from Schechter function.
%However, reionization history is not significantly different for both cases. This implies that brighter dropout galaxies have insignificant contribution in reionization process. 

%From this study it is clear that open question remain regarding the true LF especially for dropout galaxies at lower magnitude and their contribution to UV luminosity density at higher redshifts. A deeper spectroscopic observations with JWST NIRSpec and HUDF will be required for further studies~\citep{Robertson:2023, Curtis-Lake:2023}. 

\section{Acknowledgement}
 All the computations in this paper are done using the HPC Nandadevi and Kamet (\url{https://hpc.imsc.res.in}) at the Institute of Mathematical Sciences, Chennai, India. The authors would like to thank Yuichi Harikane for providing the UV luminosity data sets used in this work. DKH would like to thank Daniela Paoletti for certain important discussions. DKH would like to acknowledge the support from the Indo-French Centre for the Promotion of Advanced Research – CEFIPRA grant no. 6704-4 and the support through the India-Italy ``RELIC - Reconstructing Early and Late events In Cosmology" mobility program. 

\appendix
\section{Modification to the mean functions due to changes in the hyperparameters}\label{sec:A}
\da{In this section we describe how GP performs over its hyperparameter space. The GP likelihood in terms of data $\textbf{y}$ for zero mean function and a given kernel covariance matrix $C(X,X)$ can be written as,}
\begin{equation}
    %2\ln\, p(f|y-\mu(x)) = -(y-\mu(x))^TC(x,x)(y-\mu(x)) - \ln \,\det C(x,x) - n \ln\, (2\pi),
    \log p(\textbf{y}|{X}) = -\frac{1}{2}\textbf{y}^T(C + \sigma_{n}^2\textbf{I})\textbf{y} - \frac{1}{2} \log |C + \sigma_{n}^2\textbf{I}| -\frac{n}{2}\log\, (2\pi)
\end{equation}\label{eq:gp_like}
\begin{figure}[h]
\includegraphics[width=0.5\textwidth]{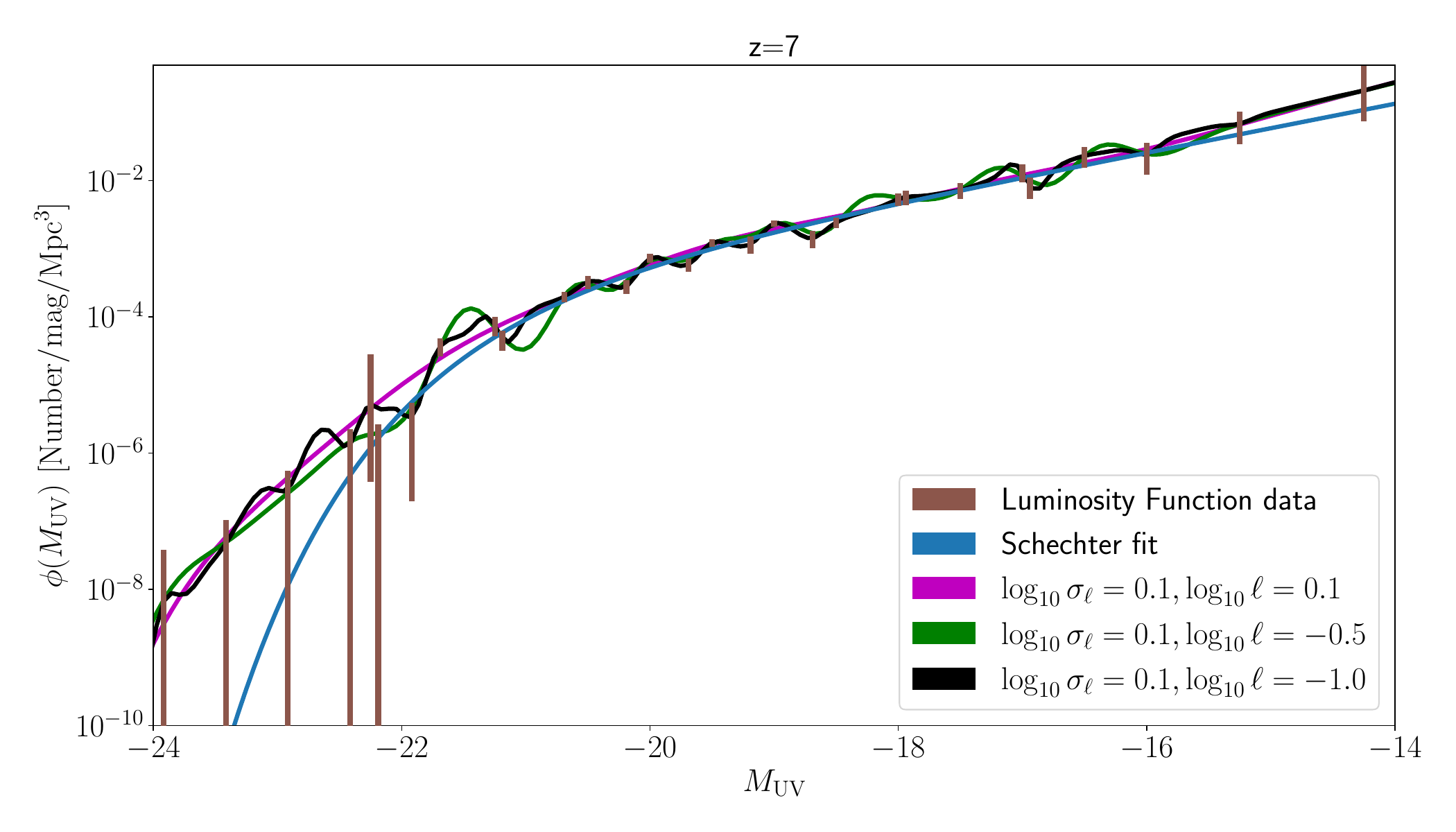}
\includegraphics[width=0.5\textwidth]{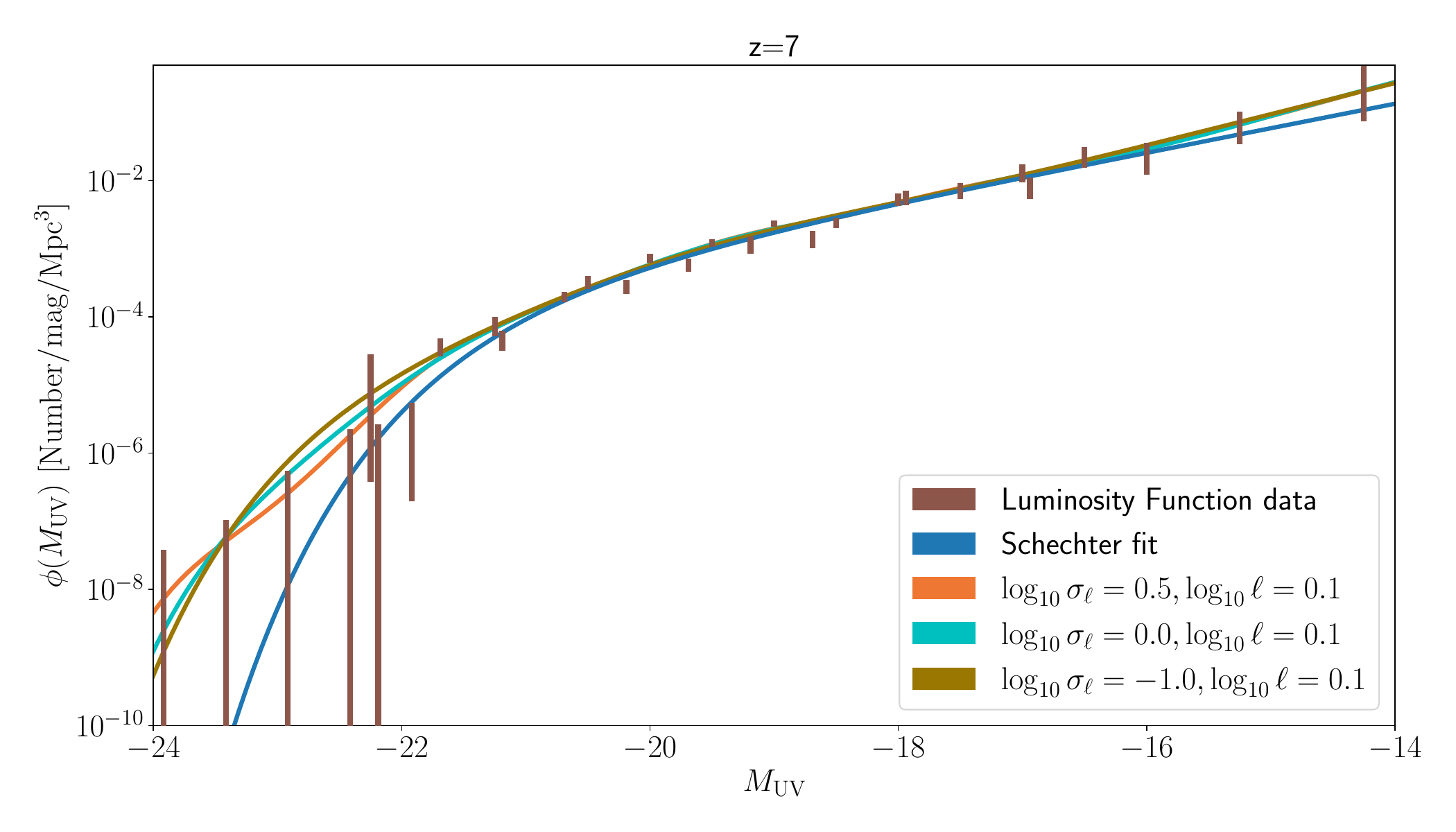}
\caption{\da{Schematic plot to show the effect of $\sigma_{\ell}$ and $\ell$ to construct LFs at redshift z$\sim$7. The blue line shows Schechter best fit. Left panel: Highlights the role of correlation length $\ell$. The black, green and magenta lines show the GP reconstructed LFs for three choice of $\log_{10}\ell = -1.0, -0.5$ and 0.1 respectively with fixed $\log_{10}\sigma_{\ell}$ = 0.1. For low value of ${\ell}$, LF is wiggly and with larger value of $\ell$ the LFs are smoother. Right panel: Highlights the role of $\sigma_{\ell}$. The dark yellow, cyan and orange lines represent the LFs for $\log_{10}\sigma_{\ell}$ = --1, 0 and 0.5 respectively for fixed $\log_{10}\ell$ = 0.1.  With different choice of $\sigma_{\ell}$, a overall sift from initial mean function is seen.}}
\label{fig:appA}
\end{figure}
\da{where $\sigma_n$ is the noise standard deviation at data points and $n$ is the number of data points. The first term is similar to the $\chi^2$ that tries to keep the GP functions as close as data points $y$ by optimizing the kernel hyperparameters. The second term is the penalty term that is independent of data $y$ and solely contributed by the kernel and this term tries to maximize the likelihood smoothing the small-scale features of GP functions by optimizing the correlation length hyperparameter $\ell$. For optimized hyperparameters, GP generates the predictive mean functions at predictive points $X_{*}$  for given mean function $\mu(X_{*})$} following,

\begin{equation}
    f({X}_{*}) = \mu(X_{*}) + \textbf{C}(X,X_{*})\textbf{C}_{y}^{-1}(\textbf{y} - \mu(X)),
\end{equation}
where $\textbf{C}_{y} = \textbf{C}(X,X) + \sigma^2_{n}\textbf{I}$.

\da{Figure~\ref{fig:appA} illustrates the role of two GP hyperparameters  $\ell$ and $\sigma_{\ell}$. We choose the data set at redshift z$\sim$ 7 as an example. In left panel, we present the role of correlation length hyperparameter $\ell$. We display the LFs from GP for given best fit Schechter mean function (blue) for fixed $\log_{10}\sigma_{\ell}$ = 0.1 and three different $\log_{10}\ell$ = 0.1 (magenta), -- 0.5 (green) and $-1.0$ (black). For low value of $\log_{10}\ell$ = --1.0, the reconstructed LF is wiggle. This choice of low correlation length is equivalent to imposing the condition to GP to prevent fitting the noise in data. With increase of correlation lengths for same $\sigma_{\ell}$, the LFs become smoother.  In right panel, we present the LFs for  $\log_{10}\sigma_{\ell}$ = --1 (dark yellow), 0.0 (cyan) and 0.1 (orange) for fixed  $\log_{10}{\ell}$ = 0.1. The curve shows an overall amplitude shift due to different choice of $\sigma_{\ell}$ which is evident below magnitude $\sim -21$ where best fit Schechter function starts sharp decay.}

%###REFERENCE ###
\bibliographystyle{JHEP}
\bibliography{ref} 
\end{document}